\documentclass[preprint,12pt,authoryear,nopreprintline]{elsarticle}
\usepackage[T1]{fontenc}
\usepackage[utf8]{inputenc}
\usepackage[letterpaper,margin=1in]{geometry}
\usepackage{times}

\usepackage{amsmath,amssymb,booktabs,array,microtype}
\usepackage{placeins,float,setspace}
\usepackage{xurl}
\usepackage[hidelinks]{hyperref}
\hypersetup{pdftitle={Accuracy Is Not Welfare: Evaluating AI Buyers under Strategic Price Complexity in a Synthetic Duopoly},pdfauthor={Jiaqu Ning},pdfsubject={Industrial organization, strategic pricing, and AI buyer welfare}}

\makeatletter
\let\ps@pprintTitle\ps@plain
\makeatother
\begin{document}
\begin{frontmatter}
\title{Accuracy Is Not Welfare: Evaluating AI Buyers under Strategic Price Complexity in a Synthetic Duopoly}
\author[inst1]{Jiaqu Ning\corref{cor1}}
\cortext[cor1]{Corresponding author}
\ead{ningjiaqu@stu.sufe.edu.cn}
\address[inst1]{Shanghai University of Finance and Economics}
\begin{abstract}
AI buying agents may alter seller incentives as well as consumer
decisions. In a synthetic duopoly, we evaluate price-comparison
pipelines on a fixed menu and in empirical correlated-equilibrium games.
Calculator assistance selects an optimal offer 84.00 percent of the time,
versus 68.57 percent under verified-total disclosure, yet disclosure has a
20.14-unit higher consumer-worst equilibrium surplus. A frozen-strategy
holdout comparison favors disclosure, while re-solving equilibria under
scenario resampling produces a range spanning zero. These model-based
results show why accuracy, loss severity, seller incentives, and welfare
incidence require separate measurement.
\end{abstract}
\begin{keyword}
AI agents \sep industrial organization \sep strategic price complexity \sep information disclosure \sep correlated equilibrium \sep consumer welfare
\JEL D18 \sep D43 \sep D83 \sep L13
\end{keyword}
\end{frontmatter}
\onehalfspacing
\raggedbottom
\hypertarget{introduction}{%
\section{Introduction}\label{introduction}}

The economic value of a decision aid depends partly on the market in
which it is used. An AI buyer that compares prices can reduce the cost
of reading contracts and calculating payments. Yet sellers have
incentives to choose the prices and descriptions to which that buyer
will respond. A procedure that performs well on a fixed collection of
offers may therefore face a different distribution of offers when it
becomes part of a market. Evaluating the procedure requires both a
measure of decision quality and a model of the incentives its decisions
create.

This problem connects AI agents to established questions in industrial
organization. Search frictions and limited information processing
influence demand; firms may then find particular prices or complex
descriptions profitable \citep{Gabaix2006, Ellison2009, Carlin2009}. Our
setting studies a computational form of this problem. Every contract
discloses its terms, and alternative presentations of a given seller's
offer have the same payment function. The difficulty is translating the
stated terms into a choice. Errors may remain even when an agent can
execute arithmetic or receives a verified total.

We ask: \textbf{How do buyer accuracy and seller incentives jointly
determine consumer welfare?} Two sellers commit to a monthly markup and
one of three fee representations before a scenario is realized. A
language-model buyer observes the contracts and service values, then
chooses one offer or an outside option. We calculate consumer surplus,
seller profit, and total surplus from the same decisions. The resulting
payoff matrices define a finite seller game; correlated-equilibrium (CE)
constraints characterize distributions under which no seller has a
profitable unilateral deviation after receiving a recommendation.

The setting is a synthetic subscription market. Its charges, usage
allowances, and conditional credits are intended to make comparisons
economically nontrivial, not to replicate a particular industry. The
model does not observe seller adjustment, estimate household
preferences, or establish the effect of a real disclosure rule. Similar
fee-comparison questions arise in payment services and retail finance,
but those markets require institutional and contractual evidence beyond
the present data. The primary contribution is to economics: it shows how
the design of an AI buyer can change the demand response entering a
strategic pricing problem.

The empirical comparison produces a ranking reversal. In 48 new core
scenarios, the calculator pipeline selects an optimal offer in 84.00\%
of decisions, versus 68.57\% for verified total-payment disclosure. With
equal weights on seller action profiles, net consumer surplus is higher
under the calculator. After solving each empirical seller game, however,
the disclosure condition has a consumer-surplus lower bound of 53.06,
compared with 32.92 for the calculator. These figures are expressed in
simulation units and depend on the specified model, costs, and action
grid.

We distinguish this full-sample equilibrium comparison from a
prespecified holdout exercise. Seller distributions are selected on 32
scenarios and evaluated without reoptimization on 16 held-out scenarios.
Disclosure's advantage is 25.51 units, with a conditional 95\% paired
scenario-bootstrap interval of {[}10.69, 42.08{]}. Re-solving
equilibrium programs under full-sample scenario resampling instead gives
a sensitivity interval of {[}-13.32, 33.77{]}. The training
distributions also admit profitable deviations under holdout payoffs.
Together, these results support a conditional fixed-strategy comparison
more clearly than they support a general claim about out-of-sample
market equilibrium.

Three features make the economic argument testable. First, the full
81-profile action grid reveals how each buyer pipeline responds to
alternative markups and formats, rather than measuring accuracy on the
observed or selected offers alone. Second, counterfactual weighting
connects the fixed-menu benchmark to the seller game. We show
algebraically that the change in the disclosure-minus-calculator
comparison equals the difference between each policy's own reweighting
term. Third, we report consumer surplus alongside seller profit and
total surplus. A consumer gain can reflect lower transfers to firms,
better allocation, or lower costs; these are distinct welfare margins.

The paper makes an empirical and methodological contribution rather than
a universal ranking of tools and disclosure. The calculator's static
advantage includes lower mean and upper-tail decision losses.
Nevertheless, its remaining errors occur in a payoff landscape that
supports different seller actions in the estimated game. Post hoc payoff
slices and cross-policy evaluations make this contrast visible without
claiming causal mediation. A randomized reasoning-budget policy performs
poorly in the later matched subset, illustrating that the distribution
of computational effort also matters. The experiment therefore motivates
economic evaluation of AI buyers as complete decision systems: accuracy,
losses, incentives, and incidence should be reported together.

The fixed-menu and strategic exercises are not alternative estimates of
a single treatment effect. They answer different questions. The former
evaluates a pipeline under a distribution chosen by the researcher,
giving equal weight to each pair of seller actions. The latter asks what
consumer outcomes remain compatible with seller incentives in games
constructed from the observed responses. The market distribution is
therefore an output of the second exercise and an input to the first.
Treating the two as interchangeable would hide the very economic channel
the design is intended to study.

This distinction matters for the interpretation of accuracy. A decision
system can be more accurate on average while making a small number of
errors at prices that are especially profitable for sellers. The profit
from a response depends on the margin on the chosen contract as well as
on the chance of being chosen. The distribution of errors over the
action grid, rather than their aggregate frequency alone, can therefore
affect the strategic comparison. Conversely, the presence of a
profitable error does not establish that a seller will find or exploit
it. That further claim would require a theory and evidence about seller
information, learning, and market institutions.

Our evaluation deliberately separates a benchmark for a buyer from a
prediction of market behavior. A uniformly weighted action grid is
useful for comparing the same set of prompts across interventions, but
it is not presented as an empirical description of existing contracts. A
correlated-equilibrium set permits flexible dependence between sellers'
recommended actions and imposes explicit unilateral incentive
constraints; it does not by itself supply a mediator or a market
selection mechanism. By presenting both exercises, the paper gives
readers the ingredients needed to assess how much of a conclusion
depends on how offers are weighted.

The contribution is also methodological. The analysis links a
reproducible AI-evaluation protocol to an economic model in which a
buyer's decision rule enters the demand faced by competing firms. It
then keeps the interpretation of each statistical object close to its
construction. A positive difference in a conditional holdout evaluation,
a sensitivity range from re-solving sampled games, and the width of a
correlated-equilibrium welfare set are not three estimates of the same
uncertainty. We report them together precisely because each leaves a
different question unresolved.

\begin{figure}[!htbp]
\centering
\includegraphics[width=\linewidth]{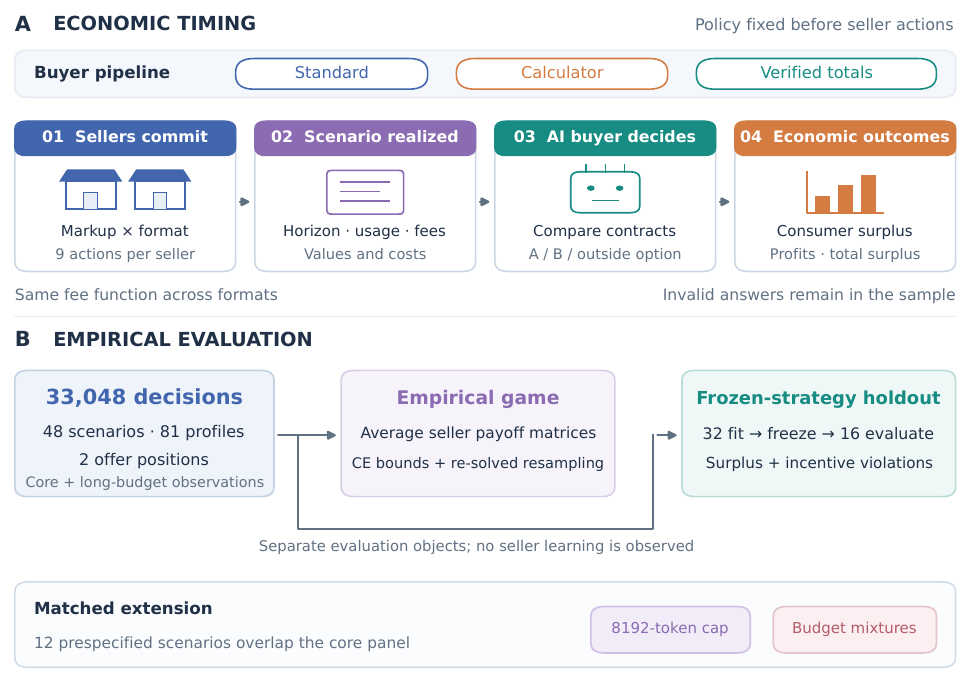}
\caption{\textbf{Economic timing and empirical evaluation.} A: sellers select markup and representation before the scenario is realized. B: measured choices support empirical CE optimization and a separate frozen-strategy holdout analysis. Static evaluation weights profiles equally. The 12-scenario extension overlaps the core; mixtures are derived payoffs. Arrows denote timing or data flow, not an observed learning process.}
\label{fig:1}
\end{figure}
\FloatBarrier

\hypertarget{related-literature}{%
\section{Related Literature}\label{related-literature}}

\hypertarget{bounded-rationality-and-strategic-complexity}{%
\subsection{Bounded rationality and strategic
complexity}\label{bounded-rationality-and-strategic-complexity}}

\citet{Simon1955} emphasizes limitations in human information
processing, while \citet{Sims2003} models scarce information capacity
within economic choice. Our token limits are engineering constraints,
not estimates of rationally chosen information acquisition, and tokens
are not equated with Shannon information. The connection is the broader
idea that processing capacity is scarce and economically consequential.

The information environment is also a strategic choice.
\citet{Gabaix2006} show how consumer myopia and shrouded attributes can
persist under competition. \citet{Ellison2009} examine search,
obfuscation, and price elasticity in an online market.
\citet{Carlin2009} places price complexity directly within a model of
retail financial competition. In our experiment, every fee term is
available, and different representations of a seller's contract produce
exactly the same payment at a given markup. The treatment concerns the
difficulty of processing complete information; it does not combine
complexity with omitted fees or fraudulent claims.

\hypertarget{disclosure-consumer-search-and-welfare}{%
\subsection{Disclosure, consumer search, and
welfare}\label{disclosure-consumer-search-and-welfare}}

Information can be available while remaining difficult to use. In
index-fund experiments, making fee information easier to find did not
always induce fee-minimizing choices \citep{Choi2010, Beshears2009}. In
a payday-lending field experiment, a more interpretable presentation of
accumulated borrowing fees changed behavior \citep{Bertrand2011}. These
are studies of financial products and human consumers; they motivate a
broad economic question about usable information rather than providing
direct evidence on our AI agents or synthetic subscription market.

The distinction also relates to price salience. \citet{Chetty2009} show
that tax presentation can affect consumer responses and welfare
analysis. Our comparison does not estimate a human salience parameter or
isolate attention from computation. Verified total payments make a
consequence explicit while the original terms remain available. The
treatment changes the information interface supplied to the buyer
pipeline, and the observed comparison includes all other differences
between that pipeline and calculator assistance.

The consumer-protection perspective emphasizes a further economic test:
whether an intervention corrects a specific decision problem at a cost
justified by its consequences \citep{Campbell2011}. Our experiment
supplies correct totals directly from the generator and omits their
verification cost. It measures choices and equilibrium payoffs
conditional on that input. A real policy assessment would require
evidence on verification, compliance, market adjustment, and the
relevant institutions.

\hypertarget{ai-economic-agents-and-strategic-evaluation}{%
\subsection{AI economic agents and strategic
evaluation}\label{ai-economic-agents-and-strategic-evaluation}}

\citet{Horton2026} discuss language models as simulated economic agents.
We study a model-operated buyer in an artificial market, without
treating its responses as measurements of household preferences.
\citet{Calvano2020} study learning by sellers in a repeated pricing
game, and \citet{Brown2023} analyze competition between sellers using
pricing algorithms. Our algorithmic intervention is on the buyer side.
Sellers are represented by estimated payoff matrices and equilibrium
optimization; we do not observe seller learning, convergence, or
algorithmic collusion. This distinction is essential when comparing the
resulting mechanisms.

Recent AI-shopping research is closely related. \citet{Buecheler2026}
use price-equivalent measures to audit seller cues across seven models,
and \citet{Wadi2026} study shopping behavior across eight models under
changes in information costs and purchasing instructions. Both are cited
as preprints. Our distinction is the combination of economically
equivalent fee representations, seller-payoff and consumer-surplus
accounting, and a new-scenario confirmation design fixed after
development. We do not rely on a claim to be the first study of
presentation effects on AI shopping.

\hypertarget{prediction-quality-decision-quality-and-incentives}{%
\subsection{Prediction quality, decision quality, and
incentives}\label{prediction-quality-decision-quality-and-incentives}}

\citet{Kleinberg2015} distinguish prediction problems from conventional
causal estimation problems and argue that improving predictive inputs
can improve decisions. Our question begins after the model's output is
placed in a decision procedure: what objective does the procedure
evaluate, and whose behavior changes in response? Predicting contractual
totals accurately is useful, but consumer welfare depends on the chosen
contract and the prices offered. Thus, the benchmark target, the
decision rule, and the market response must be specified separately.

The economics of inference also contains incentive problems on the
supply side of AI itself. \citet{Saig2024} study statistical contracts
for costly text generation, linking automated evaluation to the
incentives of a model provider. Here the strategic actors sell the
underlying products, while the buyer pipeline is held fixed within each
condition. The connection is the need to evaluate computational
performance together with the incentives attached to it. We do not
estimate an optimal contract for the AI provider, and our token-cost
parameter is not a model of provider effort or hidden production cost.

This distinction clarifies the contribution relative to broader agent
benchmarks. A task-success score aggregates decisions according to the
evaluator's menu. An economic assessment must additionally state the
losses attached to mistakes, the alternative choices available, and the
response of strategic counterparties. The relevant gap is not that
accuracy is uninformative. Accuracy is informative about a particular
event. The gap is that this event does not identify the welfare
consequences of the complete system. Our experimental records allow
those consequences to be calculated under explicit, contestable
assumptions.

\hypertarget{strategic-feedback-and-performative-evaluation}{%
\subsection{Strategic feedback and performative
evaluation}\label{strategic-feedback-and-performative-evaluation}}

The relationship between an algorithm and its input distribution has
also been studied in computer science. \citet{Hardt2016} analyze
classification when agents can adapt their observable attributes to a
deployed decision rule. \citet{Perdomo2020} study prediction when a
deployed model influences the distribution on which its performance is
evaluated. These contributions clarify why a fixed test distribution can
be an incomplete description of deployment. They do not, however, make
every market response a prediction problem, and a model of strategic
classification does not automatically determine the payoffs of competing
sellers.

Our setting connects the feedback idea to a specific economic mechanism.
The buyer pipeline is announced and held fixed within a condition;
sellers' candidate actions specify the prices and representations that
the pipeline will encounter. The subsequent joint distribution is
restricted by each seller's profits. Unlike a standard classification
task, the output is a choice among two seller-specific contracts and an
outside option. The relevant outcomes include the buyer's service value,
contractual payment, seller resource cost, and computational or
representation costs. These variables permit a welfare calculation
conditional on the artificial primitives.

The comparison with performative prediction should be kept at the level
of a shared concern about endogeneity. We do not observe a deployed
prediction system repeatedly retrained on its own induced data, and we
do not estimate the distribution map or contraction conditions in that
literature. The equilibrium program is a static solution concept applied
to measured response surfaces. It is neither an observed sequence of
model updates nor evidence that a retraining process converges. A
genuinely dynamic extension would have to specify how buyers update
their model, how sellers learn about the updates, and what information
each side observes.

Strategic classification also highlights an asymmetry in feasible
manipulation. A seller may control wording, order, and a price, but it
need not control a buyer's valuation or usage. In this experiment we
state a finite action set and hold the other primitives fixed by
scenario. That restriction is what makes a complete payoff matrix
feasible. In deployment, the feasible manipulation set is an empirical
question: product descriptions, search rankings, advertising, contract
length, quality, or bundled services could all change. Additional levers
could either amplify or weaken the specific incentives measured here.

\hypertarget{equilibrium-sets-and-the-distribution-of-surplus}{%
\subsection{Equilibrium sets and the distribution of
surplus}\label{equilibrium-sets-and-the-distribution-of-surplus}}

The game-theoretic contribution is an evaluation of welfare over a set
of incentive-consistent joint distributions, following the
correlated-equilibrium concept of \citet{Aumann1974}. A correlated
equilibrium can represent a range of patterns of dependence between
sellers' actions. This flexibility is useful for a conservative bound
because it avoids selecting a single pure or mixed Nash equilibrium in
advance. The price of that flexibility is interpretive: the existence of
a feasible joint distribution is not evidence that a market institution
implements it.

The lower bound is a welfare objective over the empirical set, not a
confidence bound. If several distributions achieve the same minimum, the
reported surplus value can be unique while the associated behavior is
not. The distinction is important when interpreting support plots: the
displayed solver-selected distribution explains one way the minimum is
attained but does not establish that every feasible market distribution
has the same action frequencies. Nor does equality between the minimum
and maximum in a fitted game remove uncertainty about the payoff matrix
from which that game was built.

The welfare comparison complements rather than replaces descriptive work
on accuracy and losses. A consumer may value lower payments even if the
corresponding change is a transfer from the seller; a social planner may
instead care about total resources, or may weight the two parties
differently. We report the components needed to make those choices
visible. Our baseline sum is not a complete social welfare function for
a real economy, because the model does not estimate marginal utilities,
externalities, public revenue, or production changes outside the two
sellers.

\hypertarget{economic-environment-and-estimands}{%
\section{Economic Environment and
Estimands}\label{economic-environment-and-estimands}}

\hypertarget{contracts-and-seller-actions}{%
\subsection{Contracts and seller
actions}\label{contracts-and-seller-actions}}

A scenario \(\theta\) specifies a contract horizon of \(T\) months and
constant monthly usage \(u\). Seller \(i\in\{1,2\}\) has a base monthly
charge \(b_i\), an excess-usage rate \(r_i\), a free allowance \(f_i\),
a monthly credit \(d_i\) available when usage reaches \(\tau_i\), and a
one-time charge \(s_i\). The seller chooses \(a_i=(m_i,k_i)\), where
monthly markup \(m_i\in\{0,4,8\}\) and representation complexity
\(k_i\in\{0,1,2\}\). Total payment is

\begin{equation}
\label{eq:1}
P_i(\theta,m_i)=s_i+T\bigl[b_i+m_i+r_i\max\{u-f_i,0\}-d_i\mathbf{1}\{u\ge\tau_i\}\bigr].
\end{equation}

The three representations state the rules directly, reorder the clauses,
or introduce annual and per-100-unit reference quantities that require
conversion. Complexity changes neither the payment function nor the
service. Shorter descriptions are padded with neutral text to match word
counts; actual token lengths are recorded separately. Each seller has
nine actions, giving 81 joint action profiles. This is the action space
of the experiment, not a demonstrated approximation to a
continuous-price equilibrium.

The buyer is given each service's full-horizon value \(v_i\) and can
choose an outside option worth zero. There is no liquidity or borrowing
constraint. Resource cost \(c_i\) is specified independently of the
retail payment. In particular, zero markup does not imply zero profit.
Contract length, usage, fees, thresholds, credits, setup charges, and
values vary across scenarios; Appendix A gives their supports.

\hypertarget{economic-outcomes}{%
\subsection{Economic outcomes}\label{economic-outcomes}}

Let \(x_i\) indicate purchase from seller \(i\), and let \(K\) be a
token-based computational-cost proxy for the full decision. Gross and
net consumer surplus, seller profit, and total surplus are

\begin{equation}
\label{eq:2}
\begin{aligned} C^G&=\sum_i x_i(v_i-P_i), & C&=C^G-\lambda K,\\ \pi_i&=x_i(P_i-c_i)-\kappa k_i, & S&=C+\pi_1+\pi_2.\end{aligned}
\end{equation}

We define \(K\) as output tokens plus 0.1 times input tokens, including
the contexts actually submitted during reasoning and final-answer
generation. The main parameters are \(\lambda=0.001\) and
\(\kappa=0.05\). They are sensitivity parameters in simulation currency
units, not estimates of GPU prices or contract-production costs.
Computation is charged even when no purchase occurs; both sellers incur
the cost of their selected representations whether or not they sell.
Invalid outputs are mapped to the outside option by the prespecified
parser and remain in every denominator.

Sellers commit to markup and representation rules before the scenario is
realized. Payoffs average over scenarios, presentation positions, and
model responses. This is an ex ante average game. A market in which
sellers observe each customer's realized scenario before selecting
prices would be a different game, and our data do not identify its
equilibrium. Figure 1 separates the economic timing from the empirical
estimation procedure.

The distinction between transfers and resource use follows directly from
the accounting. Substituting seller profits and consumer surplus into
total surplus gives

\begin{equation}
\label{eq:7}
S=\sum_i x_i(v_i-c_i)-\lambda K-\kappa(k_1+k_2).
\end{equation}

Contractual payments cancel because they are transfers from the buyer to
sellers. Service values and resource costs determine the allocation
component, while computation and representation costs reduce the
remaining surplus. This identity is useful for interpreting disclosure:
lower payments can benefit consumers without generating an equally large
social resource saving.

For a fixed allocation, a monthly markup change of \(\Delta m_i\)
changes the seller's full-horizon payment by \(T\Delta m_i\). Holding
computation and representation fixed, the buyer's surplus changes by
\(-x_iT\Delta m_i\) and that seller's profit changes by the opposite
amount. Total surplus is unchanged. When choices, computation, or
representations respond, total surplus can also move. The empirical
comparison therefore reports consumer surplus, profit, and total surplus
together; it does not assign the same economic interpretation to every
increase in consumer surplus.

These calculations adopt quasilinear utility and a common simulation
currency. They do not estimate a household's marginal utility of income.
Because the model is given service valuations, it need not infer
preferences from human behavior. This makes surplus calculations
auditable within the experiment, but the welfare numbers depend on the
investigator-specified values and costs.

\hypertarget{accuracy-is-not-welfare}{%
\subsection{Accuracy is not welfare}\label{accuracy-is-not-welfare}}

Let \(V^*(\theta,a)\) be the maximum available gross surplus and define
decision loss as \(L=V^*-C^G\). Optimal-choice accuracy records whether
\(L=0\), whereas welfare depends on the magnitude of \(L\) when it is
positive. Consider a theoretical menu with net values 100, 99, and 0. A
buyer that chooses 100 with probability 0.9 and 0 otherwise earns 90 on
average. A buyer that chooses 100 with probability 0.8 and 99 otherwise
earns 99.8. The second buyer is less accurate but delivers higher
expected surplus. This is an illustrative counterexample, not a measured
experimental result.

\textbf{Proposition 1 (representation invariance under exact
comparison).} Suppose the buyer computes true payments exactly,
maximizes gross surplus, and uses a tie-breaking rule invariant to
presentation. At a fixed markup, purchase probabilities are independent
of complexity. If \(\kappa>0\), every action with \(k_i>0\) is strictly
dominated by the action with the same markup and \(k_i=0\).
Consequently, no CE assigns positive probability to positive-complexity
actions.

\emph{Proof.} Economic equivalence implies that changing only \(k_i\)
leaves every feasible surplus and hence every purchase probability
unchanged, for every opponent action. Replacing positive complexity with
zero complexity leaves expected revenue and production cost unchanged
but reduces the seller's cost by \(\kappa k_i>0\). Such an action is
strictly dominated and cannot have positive probability in a correlated
equilibrium. \(\square\)

This proposition describes an ideal benchmark, not the language model in
the disclosure condition. Even with correct totals available, the model
must compare values and choose an option. It can ignore information,
miscompare utilities, or fail after truncation. Residual complexity
effects in the experiment therefore do not contradict the benchmark.

\hypertarget{empirical-equilibrium-bounds-and-holdout-performance}{%
\subsection{Empirical equilibrium bounds and holdout
performance}\label{empirical-equilibrium-bounds-and-holdout-performance}}

For each intervention \(h\), we estimate payoff matrices by averaging
equally over scenarios and the two offer positions. Following
\citet{Aumann1974}, a joint action distribution \(q\) is a CE when no
seller benefits from deviating after receiving an action recommendation:

\begin{equation}
\label{eq:3}
\sum_{a_{-i}}q(a_i,a_{-i})\bigl[\bar\pi_i^h(a_i,a_{-i})-\bar\pi_i^h(a_i^{\prime},a_{-i})\bigr]\ge0,\quad\forall i,a_i,a_i^{\prime}.
\end{equation}

These are joint-probability-weighted incentive constraints, rather than
conditional deviations divided by the probability of each
recommendation. Linear programs yield the lowest and highest net
consumer surplus within the empirical CE set:

\begin{equation}
\label{eq:4}
\begin{aligned}\underline C^h&=\min_{q\in\mathrm{CE}(\bar\pi^h)}\sum_a q(a)\bar C^h(a),\\\overline C^h&=\max_{q\in\mathrm{CE}(\bar\pi^h)}\sum_a q(a)\bar C^h(a).\end{aligned}
\end{equation}

The bounds describe the equilibrium set of the estimated game. They
neither predict equilibrium selection in an actual market nor assert the
presence of a coordination device. We use the consumer-worst CE as a
conservative evaluation criterion and save the solver's selected
distribution whenever an optimum is not unique.

To relate the strategic comparison to the fixed-menu benchmark, let
\(q^U\) assign equal probability to the 81 joint actions. Write
\(W^h(q)\) for mean net consumer surplus under buyer pipeline \(h\) and
joint action distribution \(q\). Let \(q^D\) and \(q^C\) be the selected
consumer-worst empirical CE distributions for disclosure and calculator
assistance. Define \(\Delta^U=W^D(q^U)-W^C(q^U)\) and
\(\Delta^{\mathrm{CE}}=W^D(q^D)-W^C(q^C)\). Then

\begin{equation}
\label{eq:10}
\begin{gathered}\Delta^{\mathrm{CE}}-\Delta^U=\bigl[W^D(q^D)-W^D(q^U)\bigr]\\-\bigl[W^C(q^C)-W^C(q^U)\bigr].\end{gathered}
\end{equation}

The left side is the change in the policy comparison when the equally
weighted menu is replaced by policy-specific incentive-consistent
distributions. Each term on the right holds the buyer pipeline fixed
while changing the weights on its own payoff surface. This identity is
accounting, not a causal mediation decomposition: the CE distributions
are estimated under different pipelines, and the uniform distribution
need not be an equilibrium. It makes the economic role of offer weights
explicit.

For out-of-sample evaluation, we solve for \(\widehat q^h\) using
training scenarios only and hold that distribution fixed on test
scenarios. The primary statistic compares disclosure \(D\) with
calculator assistance \(C\):

\begin{equation}
\label{eq:5}
\widehat\Delta_{\mathrm{test}}=\frac{1}{n_{\mathrm{test}}}\sum_{\theta\in\mathcal T}\bigl[C_\theta^D(\widehat q^D)-C_\theta^C(\widehat q^C)\bigr].
\end{equation}

Here \(\mathcal T\) is the holdout set and \(C_\theta^h(q)\) is the
position-averaged net surplus under joint distribution \(q\). This
estimand is conditional on the fitted training distributions; it is not
a difference between newly estimated holdout equilibria. We also report
the maximum joint deviation gain of each frozen distribution under
holdout-average seller payoffs. A positive gain means that the
distribution is not an exact CE of that holdout game.

To formalize the incentive diagnostic, define the largest joint
deviation gain as

\begin{equation}
\label{eq:8}
\begin{gathered}G_i^h(a_i,a_i^{\prime};q)=\sum_{a_{-i}}q(a_i,a_{-i})\bigl[\bar\pi_i^h(a_i^{\prime},a_{-i})-\bar\pi_i^h(a_i,a_{-i})\bigr],\\\varepsilon^h(q)=\max\bigl\{0,\max_{i,a_i,a_i^{\prime}}G_i^h(a_i,a_i^{\prime};q)\bigr\}.\end{gathered}
\end{equation}

The outer maximum includes zero. The statistic is zero exactly when the
displayed CE inequalities hold, up to numerical tolerance. Its units are
expected simulation profit, and the gain is multiplied by the
probability of receiving the corresponding recommendation. Dividing a
particular constraint by a very small recommendation probability would
produce a different, conditional measure. We retain the joint form
because it is the object constrained by the linear program and avoid
interpreting its numerical value as a bound on every conditional
temptation to deviate.

A distribution may yield favorable holdout consumer surplus while
violating holdout seller incentives. Conversely, a distribution can
satisfy the incentive constraints and still leave consumers with low
surplus. Reporting these objects together prevents a common interpretive
shortcut: neither high welfare nor low deviation gain alone establishes
that an intervention has produced a desirable, stable market outcome.
The outcome and stability criteria concern different parts of the
evaluation.

We do not assume that a mediator would implement the computed
distribution or that sellers know the complete payoff matrix. CE
supplies a tractable set of incentive-consistent distributions for the
estimated finite game. The experiment describes that set and examines
one selected extreme point. Strategic learning dynamics, the choice
among multiple payoff-equivalent distributions, and robustness to agents
with different information remain separate empirical questions.

\hypertarget{demand-responses-and-the-profitability-of-complexity}{%
\subsection{Demand responses and the profitability of
complexity}\label{demand-responses-and-the-profitability-of-complexity}}

A seller's incentive to alter a representation depends on how the change
affects the chance of purchase at each set of contractual primitives.
Holding its markup and the rival's action fixed, let \(x_i(k)\) denote
the purchase indicator under representation \(k\) and write the
expectation over the scenario and position distribution used to
construct the game. The net-profit change from replacing the direct
format by a more complex format is the expectation of
\((P_i-c_i)[x_i(k)-x_i(0)]\) minus \(\kappa k\). Payment and production
cost here are evaluated for the same underlying offer. This identity
states the economic margin without assuming that the probability change
has the same sign in every scenario.

If the additional format attracts purchases in high-margin situations
but loses them in low-margin situations, the profit effect can be
positive even when its average effect on the purchase rate is small. The
reverse can also occur. Thus a scalar average accuracy, or even an
average change in demand, does not generally recover the strategic
incentive to change a format. The identity also explains why an
apparently small representation cost does not determine whether the
format is chosen: the cost must be compared with the margin-weighted
response over the relevant opponent actions and scenarios.

This is an accounting relationship, not an estimated elasticity. Our
action grid contains three discrete markups and three designed
representations, so the reported payoff slices compare finite
alternatives. The labels 0, 1, and 2 identify particular templates and
do not establish a cardinal scale on which one unit is the same amount
of complexity across products or languages. Since a template changes
wording and structure together, the experiment does not isolate clause
order, numerical conversion, or the position of any single fee term. A
component-level intervention would be needed for those conclusions.

The same qualification applies to markup incentives. Moving from one
price in the grid to another changes the payment conditional on purchase
and can change the probability and identity of the purchase. An increase
in markup may raise revenue per sale while reducing demand; with an
imperfect AI buyer, it may instead attract sales in some scenarios. The
economic question is therefore not whether an error exists, but whether
the pattern of errors changes the relative profitability of feasible
actions. Figure 6 reports one slice of this relationship while keeping
the rival's action fixed; it is not a complete best-response map against
every possible mixed strategy.

The ex ante timing is also central. Sellers select rules before the
scenario is realized, and their payoff matrices average across the
specified scenario distribution. Consequently, their action is not a
separate personalized price for every realized buyer profile. If sellers
observed usage and valuation before choosing a contract, the relevant
incentive constraints would condition on a richer information set. The
present results should not be read as an estimate of personalized
pricing or as evidence about price discrimination against particular
users.

\hypertarget{distributional-reweighting-and-verification-costs}{%
\subsection{Distributional reweighting and verification
costs}\label{distributional-reweighting-and-verification-costs}}

The accounting identity in Section 3.4 can be complemented by a simple
bound. Let \(q\) and \(q'\) be any two distributions over the finite
action grid, and suppose the absolute value of a policy's
scenario-averaged net consumer surplus is bounded by \(M\) at every
joint action. Define total variation as one half of the sum of the
absolute differences between the two distributions. Then the absolute
difference between \(W^h(q)\) and \(W^h(q')\) cannot exceed \(2M\) times
that total variation. This follows by applying the triangle inequality
to the action-by-action weighted sum.

The bound is deliberately general. It does not require either
distribution to be an equilibrium and does not explain why the weights
differ. It shows that conclusions based on a fixed action distribution
can be stable when both the distribution shift and the range of possible
payoffs are small. A large shift can matter only if the payoff surface
also varies across the actions receiving different weights. In our
experiment the post hoc reweighting table reports the realized change in
the weighted averages; we do not estimate this general bound as an
additional experimental outcome.

An analogous calculation clarifies the unmeasured cost of producing
verified totals. Suppose, purely as a counterfactual, that generating
and maintaining disclosure information adds a fixed cost \(F\) shared
evenly across \(n\) decisions, that this cost is charged to the
consumer, and that it does not change either seller's revenue or
action-dependent cost. All disclosure consumer payoffs then fall by
\(F/n\), while the seller payoff matrices and their
correlated-equilibrium sets remain unchanged. The lower bound on
disclosure consumer surplus also falls by exactly \(F/n\).

Under those strong assumptions, the full-sample difference of 20.14
simulation units would be eliminated at \(F/n=20.14\), before accounting
for uncertainty in the measured game. This is a break-even identity, not
an estimate of the real cost or number of users required for a
disclosure system. If verification costs vary with contract complexity,
are paid by sellers, affect entry, or alter the buyer's decision
process, the seller game itself can change and a simple subtraction is
insufficient. The existing experiment measures none of these production
channels.

A different incidence assumption changes the welfare object. If a third
party pays the cost without passing it to consumers, measured consumer
surplus may remain unchanged while aggregate surplus falls. If the fee
is financed through taxes or a platform charge, a distributional
analysis must state who pays and whether financing changes behavior. The
arithmetic is straightforward once the incidence and behavioral
assumptions are explicit; the existing data do not determine those
assumptions.

\hypertarget{experimental-design}{%
\section{Experimental Design}\label{experimental-design}}

\hypertarget{development-and-confirmation}{%
\subsection{Development and
confirmation}\label{development-and-confirmation}}

The project has three stages: an initial pilot, protocol diagnostics and
calibration, and confirmation on new scenarios. Early observations
revealed frequent reasoning-budget exhaustion and answer-protocol
problems. Development diagnostics informed the 8,192-token high-budget
cap, while separate calibration scenarios fixed the probability used in
cost-matched budget randomization. The primary comparison, scenario set,
and split rule were then fixed for confirmation. This is a
development-informed confirmation design, not a third-party
preregistration.

\begin{table}[!htbp]
\caption{Experimental stages and the role of each sample.}
\label{tab:1}
\centering\small
\setlength{\tabcolsep}{3.2pt}
\renewcommand{\arraystretch}{1.12}
\begin{tabular}{@{}>{\raggedright\arraybackslash}p{98.64pt}>{\centering\arraybackslash}p{51.46pt}>{\raggedright\arraybackslash}p{184.42pt}>{\raggedright\arraybackslash}p{94.35pt}@{}}
\toprule
\textbf{Stage} & \textbf{Decisions} & \textbf{Design} & \textbf{Use} \\ \midrule
Initial pilot & 2,268 & 4 main + 4 development scenarios & Exploration \\
Diagnostics / calibration & 12,456 & 8B diagnostics and calibration; 14B development & Protocol selection \\
Core confirmation & 31,104 & 48 scenarios $\times$ 81 profiles $\times$ 2 positions $\times$ 4 conditions & Primary evidence \\
Long-budget confirmation & 1,944 & 12 scenarios $\times$ 81 profiles $\times$ 2 positions & Prespecified subset \\
\bottomrule
\end{tabular}
\vspace{3pt}\par\begin{minipage}{\linewidth}\footnotesize The final two rows total 33,048 new decisions. Long-budget scenarios overlap the core panel. Decision counts are not independent sample sizes.\end{minipage}
\end{table}

The stages contain 47,772 decisions in total, but their protocols differ
and development results had already been examined. They cannot be pooled
into one confirmatory sample. Core confirmation uses scenario seeds
10000 through 10047. A scenario is held out when
\((\mathrm{seed}-10000)\bmod 3=2\), producing 32 training and 16 holdout
scenarios. The long-budget subset takes every fourth seed starting at
10000, yielding 12 scenarios and an 8/4 split. It was specified in
advance, overlaps the core panel, and is not an independent replication.

\hypertarget{interventions-and-generation-protocol}{%
\subsection{Interventions and generation
protocol}\label{interventions-and-generation-protocol}}

The standard buyer reads the complete contracts and returns A, B, or N.
The calculator buyer generates arithmetic expressions for the two total
payments; a restricted arithmetic interpreter evaluates them, and a
deterministic routine compares utilities. The disclosure buyer receives
the full contracts plus both verified total payments and still makes the
final choice itself. Thus, calculator assistance and disclosure are
different complete pipelines. Their comparison cannot isolate the effect
of a single prompt component or of adding a tool while holding every
other feature constant.

\begin{table}[!htbp]
\caption{Interventions and the location of computational assistance.}
\label{tab:2}
\centering\small
\setlength{\tabcolsep}{3.2pt}
\renewcommand{\arraystretch}{1.12}
\begin{tabular}{@{}>{\raggedright\arraybackslash}p{77.20pt}>{\raggedright\arraybackslash}p{171.55pt}>{\centering\arraybackslash}p{68.62pt}>{\raggedright\arraybackslash}p{111.51pt}@{}}
\toprule
\textbf{Pipeline} & \textbf{Additional input or operation} & \textbf{Reasoning cap} & \textbf{Final decision} \\ \midrule
Standard & Complete original contracts & 256 / 1024 / 8192 & Model selects A, B or N \\
Calculator & Model writes two price expressions; interpreter evaluates & 256 & Deterministic utility comparison \\
Disclosure & Complete contracts plus verified total payments & 256 & Model selects A, B or N \\
Random budget & Mixture of standard endpoint payoff matrices & 256 or 8192 & Endpoint procedure \\
\bottomrule
\end{tabular}
\vspace{3pt}\par\begin{minipage}{\linewidth}\footnotesize All pipelines retain given service values and the outside option. The long cap is measured on 12 scenarios. Random policies are derived mixtures. Final-answer generation has a separate 256-token cap.\end{minipage}
\end{table}

The distinction in Table 2 is consequential. The calculator pipeline
must first translate natural-language rules into expressions, after
which arithmetic execution and utility comparison are deterministic. The
disclosure pipeline receives correct totals but retains a model-based
final comparison. A valid calculator expression can omit a charge or
implement a credit incorrectly; a valid disclosure response can still
choose the worse option. Each pipeline therefore has a different
location at which an error can enter the decision.

The disclosure treatment also does not supply a recommendation. Both
totals may be high, and differences in given service values can make the
cheaper offer suboptimal. Supplying verified totals simplifies the
payment component without removing the need to compare utilities or
consider the outside option. This design helps distinguish an
informational input from an oracle that directly supplies the correct
action.

Confirmation uses a pinned Qwen3-8B checkpoint \citep{Yang2025}, BF16,
and vLLM 0.11.0 \citep{Kwon2023}. Reasoning uses temperature 0.6,
top-\(p\) 0.95, and top-\(k\) 20. When the reasoning cap is reached, a
fixed transition prompt requests the answer; final JSON generation uses
temperature zero and a cap of 256 tokens. A reasoning budget is
therefore not the full token consumption of a decision. Each scenario,
action profile, and condition is evaluated once in each of two swapped
offer positions. Shared seed rules pair the runs but do not guarantee
identical stochastic paths across different prompts.

All contracts and response instructions are in English. Templates,
padding rules, model revision, and scenario seeds are fixed. The 33,048
confirmation decisions ran on one A100-SXM4-80GB. Recorded inference
took 6.229 hours; the worker, including model initialization, took 6.281
hours. These figures do not include every installation or download cost.
There are no additional measured H100 results in the evidence used here.

\hypertarget{uncertainty-and-interpretation}{%
\subsection{Uncertainty and
interpretation}\label{uncertainty-and-interpretation}}

Each core condition has \(48\times81\times2=7,776\) decisions. The
scenario, not the individual model call, is the resampling unit. We
resample all actions, positions, and interventions within a selected
scenario together. The primary interval is a percentile interval from
1,000 paired bootstraps of the 16 holdout scenario differences,
conditional on the training distributions. Secondary comparisons are not
multiplicity-adjusted and are not treated as co-primary tests.

We separately resample complete empirical games 1,000 times, re-solving
the CE programs after every draw, and perform leave-one-scenario-out
analyses. Linear-program optima and supports can change non-smoothly
with estimated payoffs. \citet{Fang2019} show why conventional bootstrap
validity for non-smooth functionals requires care. We establish no
coverage theorem for this CE mapping and therefore call its re-solved
percentile ranges empirical sensitivity intervals.

The design measures differences among specified computational pipelines
in a fixed synthetic environment. It does not identify real consumer
demand elasticities, firms' actual pricing responses, or the causal
effect of a disclosure regulation. Moreover, one generation per position
leaves model-sampling noise within each payoff cell. Scenario resampling
does not replace repeated independent draws from the same prompt.

\hypertarget{units-of-analysis-and-diagnostic-discipline}{%
\subsection{Units of analysis and diagnostic
discipline}\label{units-of-analysis-and-diagnostic-discipline}}

The experiment contains many decisions because every scenario is crossed
with a complete action grid and both offer positions. This is coverage
of an economic response surface, not 33,048 independent draws from a
market population. Within a scenario, contracts share horizons, usage,
fees, and values. Responses across profiles consequently inherit common
economic difficulty. Pairing conditions on the same scenarios removes
changes in those primitives from the comparison, while preserving their
dependence in uncertainty calculations.

We distinguish three uses of the observations. Static summaries average
over the evaluator's complete menu. Strategic summaries first aggregate
observations into payoff matrices and then solve an
incentive-constrained optimization problem. Holdout summaries fit a
distribution on one set of scenarios and evaluate that already selected
distribution on another set. The resulting averages can differ
substantially even when they originate from the same model calls.
Interpreting them requires their weighting scheme, not just the number
of rows used to calculate them.

For the expanded diagnostics, the canonical batch records are read
directly from the hash-verified confirmation archive. The duplicated
JSONL export is excluded. Response-category counts, loss percentiles,
and fixed-opponent profit slices are calculated without drawing a new
scenario or generating a new response. They are identified as post hoc
in the underlying tables and captions. No additional hypothesis test is
attached to these descriptions, and they do not change the prespecified
primary comparison.

The report also distinguishes nontrading outcomes from technical
failures. Choosing the outside option can be optimal when both offers
have negative net value. A failed answer is mechanically mapped to that
same option, so an invalid output may occasionally be economically
correct. Treating every nontrade as failure would misclassify rational
abstention; dropping invalid answers would change the estimand and
selectively remove costly mistakes. Keeping both indicators makes this
distinction visible.

\hypertarget{why-the-action-grid-and-scenario-split-are-distinct}{%
\subsection{Why the action grid and scenario split are
distinct}\label{why-the-action-grid-and-scenario-split-are-distinct}}

The experimental design varies two different objects: the underlying
contract scenario and the strategic actions available to sellers. A
scenario supplies values, usage, and fee primitives. Conditional on
those primitives, each seller can choose among nine markup-format pairs.
Crossing the two sellers' choices generates the 81 joint profiles
evaluated for each buyer condition and offer order. This structure
allows a payoff matrix to be built without assuming that the seller
chooses only a single observed action.

The split holds out scenarios, not individual cells of the action
matrix. This matters because two actions evaluated in the same scenario
share the same horizon, usage, and underlying service environment.
Splitting individual calls at random would put closely related versions
of the same scenario in both training and test samples. A scenario-level
split gives a more meaningful check of whether the fitted seller
distribution performs on a new draw of contract primitives. It is still
a split within the same synthetic generator, and successful transfer
within that generator is not evidence of transfer to a different
industry or population.

The two position orders serve a separate purpose. When the same two
economic offers are shown in reversed positions, any order sensitivity
can affect which seller is chosen and thus each seller's measured
profit. Averaging the two positions reduces dependence on one arbitrary
labeling convention. It does not prove that order effects vanish, since
there is one stochastic generation for each order and prompt. Nor does
swapping positions generate an additional independent scenario. Both
observations remain clustered with the scenario and joint action that
produced them.

The 48-scenario core is designed for complete coverage of the chosen
finite action set rather than for a large-sample estimate of a
population of firms. This prioritizes comparability of payoff matrices
across buyer conditions and makes the strategic computation transparent.
Its limitations follow from the same choice: the grid cannot reveal what
happens between tested markups, under other contract representations, or
after an untested change to the buyer. A larger grid would improve
coverage but also multiply the number of prompts and increase the
difficulty of estimating each payoff precisely.

Training, evaluation, and reoptimization must therefore be kept
separate. The training equilibrium is obtained from averages on 32
scenarios. Evaluating that fixed distribution on 16 scenarios changes
the scenario average but does not allow the distribution to adapt.
Re-solving on a bootstrap sample changes both the estimated game and,
potentially, the selected optimizer. Neither procedure is intrinsically
the one correct analysis for every question: the first addresses
transfer of a fixed object; the second exposes sensitivity to changes in
estimated payoffs. Their intervals condition on different operations and
cannot be compared as if they were interchangeable standard errors.

\hypertarget{the-unit-of-treatment-is-a-complete-buyer-pipeline}{%
\subsection{The unit of treatment is a complete buyer
pipeline}\label{the-unit-of-treatment-is-a-complete-buyer-pipeline}}

An intervention in this study is defined by the entire path from
information to action. The standard condition delegates reading,
computation, and choice to the language model. The calculator condition
delegates the translation of contract terms into arithmetic expressions
to the model, but uses a restricted interpreter and deterministic
utility comparison afterward. The disclosure condition supplies correct
total payments and retains model-based comparison and choice. These
arrangements differ in more than one component, so the observed
difference is a pipeline comparison.

This distinction prevents an attractive but unsupported causal story. A
difference between calculator and disclosure cannot be assigned solely
to the addition of arithmetic execution, solely to the presence of
verified totals, or solely to a change in the prompt. Those elements are
bundled with a different final decision rule. The design does establish
the performance of the specified bundles on the observed menu. A
factorial experiment would be needed to isolate the separate
contributions of information, expression generation, execution, and
final choice while holding the remaining components constant.

The calculator also illustrates why nominal access to a tool is not the
same as successful use. The interpreter can faithfully evaluate a valid
expression that does not represent the contract. A system may therefore
be syntactically valid and economically wrong. Likewise, if expression
parsing fails and the prescribed fallback is no trade, a decision can be
optimal by chance when the outside option is genuinely best. We preserve
these cases in the outcome rather than converting them into corrected
decisions or silently removing them from the sample.

Real implementations may add retries, independent extraction, confidence
checks, or human review. Such additions could change both the error
distribution and the cost of the pipeline. They might also change the
incentives in the seller game by moving errors away from profitable
price-format combinations. The present analysis cannot infer their
performance by retrospectively correcting existing outputs. Each new
workflow would require its own responses on the relevant action grid
and, for a strategic comparison, a new payoff matrix.

The same care applies to the verified totals. They are correct because
the scenario generator knows the contracts. An actual disclosure
provider would have to obtain up-to-date terms, interpret conditions,
handle errors, and decide how to resolve disputes. Treating those inputs
as free and errorless defines an informative benchmark; it does not
establish the feasibility, reliability, or net benefit of a particular
disclosure institution. We return to these missing production and
governance costs when discussing policy.

\hypertarget{results}{%
\section{Results}\label{results}}

\hypertarget{a-reversal-between-static-accuracy-and-strategic-welfare}{%
\subsection{A reversal between static accuracy and strategic
welfare}\label{a-reversal-between-static-accuracy-and-strategic-welfare}}

Table 3 and Figure 2 use the same 48 scenarios. Standard reasoning at
budget 256 achieves 32.99\% optimal choices and average gross consumer
surplus of -2.39. Increasing the cap to 1,024 raises these figures to
57.92\% and 15.30. The calculator reaches 84.00\% and 31.90, compared
with 68.57\% and 26.89 for disclosure. The disclosure-minus-calculator
accuracy difference is -15.43 percentage points, with a paired scenario
interval of {[}-23.01, -7.79{]}. The gross-surplus difference is -5.01,
with interval {[}-9.62, -0.95{]}. Both static criteria favor the
calculator.

\begin{table}[!htbp]
\caption{Static choice performance and strategic welfare on the core panel.}
\label{tab:3}
\centering\small
\setlength{\tabcolsep}{3.2pt}
\renewcommand{\arraystretch}{1.12}
\begin{tabular}{@{}>{\raggedright\arraybackslash}p{104.02pt}>{\centering\arraybackslash}p{62.41pt}>{\centering\arraybackslash}p{62.41pt}>{\centering\arraybackslash}p{62.41pt}>{\centering\arraybackslash}p{62.41pt}>{\centering\arraybackslash}p{62.41pt}@{}}
\toprule
\textbf{Condition} & \textbf{Optimal (\%)} & \textbf{Gross CS} & \textbf{Net CS lower} & \textbf{Seller profit} & \textbf{Total surplus} \\ \midrule
Standard 256 & 32.99 & -2.39 & -31.00 & 89.75 & 58.75 \\
Standard 1024 & 57.92 & 15.30 & -1.12 & 53.29 & 52.17 \\
Calculator 256 & 84.00 & 31.90 & 32.92 & 49.43 & 82.34 \\
Disclosure 256 & 68.57 & 26.89 & 53.06 & 34.92 & 87.98 \\
\bottomrule
\end{tabular}
\vspace{3pt}\par\begin{minipage}{\linewidth}\footnotesize 48 scenarios. Accuracy and gross CS equally weight profiles and positions. The final three columns use each intervention’s consumer-worst empirical CE, with $\lambda$ = 0.001 and $\kappa$ = 0.05. Monetary outcomes are simulation units.\end{minipage}
\end{table}

\begin{figure}[!htbp]
\centering
\includegraphics[width=\linewidth]{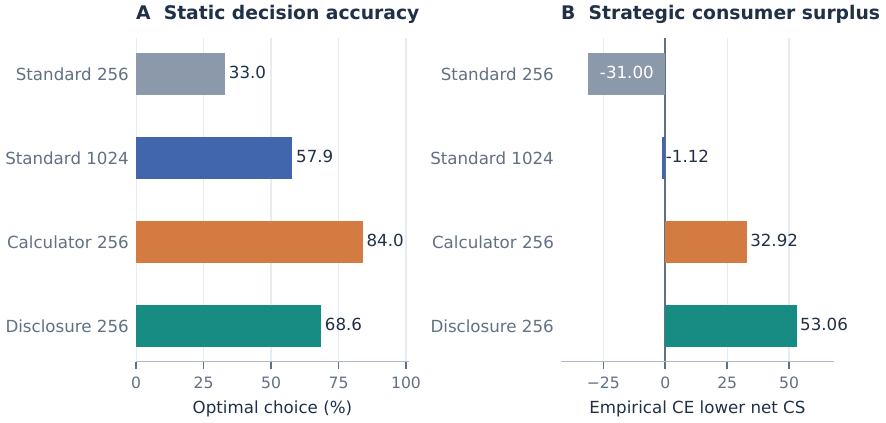}
\caption{\textbf{The same decisions imply different static and strategic rankings.} Both panels use all 48 core scenarios. Static accuracy weights profiles and positions equally; strategic welfare uses each empirical CE lower bound. $\lambda$ = 0.001 and $\kappa$ = 0.05. No uncertainty is encoded by bar lengths.}
\label{fig:2}
\end{figure}

Mapping these same choices into seller profits changes the ordering. The
empirical CE lower bound on net consumer surplus is 53.06 under
disclosure and 32.92 under the calculator, a difference of 20.14. Both
exceed the -1.12 value under standard reasoning at 1,024. At the main
cost parameters, the lower and upper consumer-surplus bounds coincide
within numerical tolerance in all four core conditions. This equality
concerns a welfare functional; it does not establish uniqueness of the
joint strategy distribution or precise identification of a population
equilibrium.

The economic distinction is the weighting of offers. Under uniform
action-profile weights, average net surplus is 31.43 for the calculator
and 26.47 for disclosure, so \(\Delta^U=-4.96\). Under the respective
consumer-worst empirical CE distributions,
\(\Delta^{\mathrm{CE}}=20.14\). Thus the accounting wedge in the
preceding identity is 25.10 units. Relative to the uniform benchmark,
the CE weighting raises the disclosure value by 26.58 and the calculator
value by 1.48 (Table 11). These are post hoc descriptions of the same
payoff matrices, not independently estimated causal effects or
additional observations.

High average accuracy does not determine which pricing and presentation
profiles retain errors, or prevent sellers from finding those profiles
profitable. The same buyer responses can therefore imply different
welfare comparisons under different offer weights. The calculation is a
precise explanation of the observed ranking reversal within the finite
empirical game; it does not establish how suppliers would adjust prices
in an actual market.

\hypertarget{the-primary-holdout-comparison}{%
\subsection{The primary holdout
comparison}\label{the-primary-holdout-comparison}}

After fixing seller distributions on the 32 training scenarios, average
net consumer surplus on the 16 holdout scenarios is 52.02 under
disclosure and 26.51 under the calculator. The prespecified difference
is 25.51, with a 95\% conditional interval of {[}10.69, 42.08{]}.
Relative to standard reasoning at 1,024, the calculator and disclosure
differences are 33.79 and 59.31, respectively (Table 4).

\begin{table}[!htbp]
\caption{Holdout net consumer-surplus differences with training distributions fixed.}
\label{tab:4}
\centering\small
\setlength{\tabcolsep}{3.2pt}
\renewcommand{\arraystretch}{1.12}
\begin{tabular}{@{}>{\raggedright\arraybackslash}p{180.13pt}>{\centering\arraybackslash}p{64.33pt}>{\centering\arraybackslash}p{124.37pt}>{\centering\arraybackslash}p{60.04pt}@{}}
\toprule
\textbf{Comparison} & \textbf{Difference} & \textbf{95\% conditional interval} & \textbf{Role} \\ \midrule
Disclosure $-$ calculator & 25.51 & [10.69, 42.08] & Primary \\
Calculator $-$ standard 1024 & 33.79 & [15.89, 54.65] & Secondary \\
Disclosure $-$ standard 1024 & 59.31 & [32.48, 90.72] & Secondary \\
\bottomrule
\end{tabular}
\vspace{3pt}\par\begin{minipage}{\linewidth}\footnotesize 32 training and 16 holdout scenarios; 1,000 paired scenario bootstraps. Intervals condition on fitted training distributions. Secondary comparisons are not multiplicity-adjusted.\end{minipage}
\end{table}

\begin{figure}[!htbp]
\centering
\includegraphics[width=\linewidth]{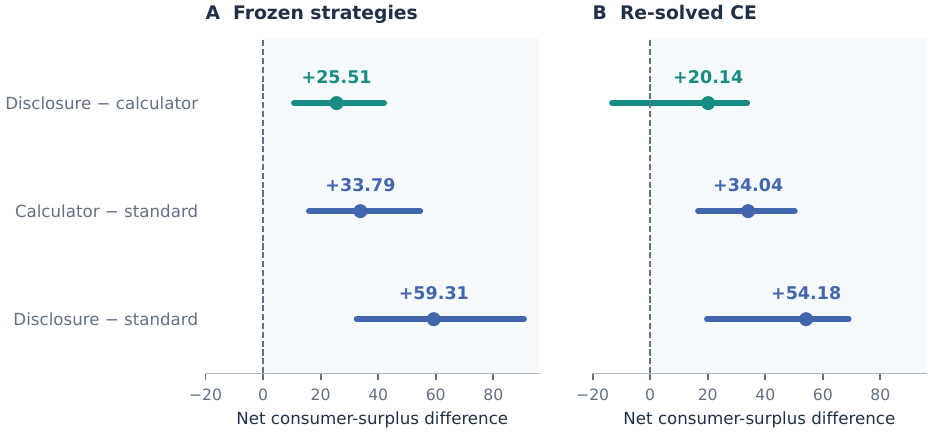}
\caption{\textbf{Conditional holdout intervals and equilibrium sensitivity answer different questions.} A: 95\% paired percentile intervals conditional on 32-scenario training distributions, evaluated on 16 holdout scenarios. B: 2.5th–97.5th percentiles after re-solving games on resampled 48-scenario panels. Each uses 1,000 draws. Standard denotes budget 1024; shading denotes positive differences.}
\label{fig:3}
\end{figure}

This result supports better performance on new scenarios for the frozen
training distributions. It does not incorporate all uncertainty from
drawing a different training sample and selecting a different
equilibrium. The maximum joint deviation gains in the holdout-average
game are 0.4633 for disclosure, 8.0675 for the calculator, and 4.9697
for standard reasoning at 1,024. None of these distributions is an exact
holdout CE. Accordingly, the primary difference should not be described
as an out-of-sample market-equilibrium treatment effect.

Table 5 and Figure 4 show how the selected distributions transfer.
Calculator net consumer surplus falls from 37.61 on training scenarios
to 26.51 on holdout scenarios. Disclosure falls from 53.11 to 52.02.
These level changes are descriptive and depend on the realized split;
the paired treatment contrast remains the primary statistic. They do not
establish that one pipeline has a generally smaller generalization error
over repeated training samples.

\begin{table}[!htbp]
\caption{Generalization of the selected training distributions.}
\label{tab:5}
\centering\small
\setlength{\tabcolsep}{3.2pt}
\renewcommand{\arraystretch}{1.12}
\begin{tabular}{@{}>{\raggedright\arraybackslash}p{122.52pt}>{\centering\arraybackslash}p{63.37pt}>{\centering\arraybackslash}p{76.05pt}>{\centering\arraybackslash}p{76.05pt}>{\centering\arraybackslash}p{84.49pt}@{}}
\toprule
\textbf{Condition} & \textbf{Support size} & \textbf{Training CS} & \textbf{Holdout CS} & \textbf{Holdout deviation} \\ \midrule
Standard 256 & 1 & -28.07 & -36.87 & 0.0000 \\
Standard 1024 & 4 & 4.13 & -7.29 & 4.9697 \\
Calculator 256 & 1 & 37.61 & 26.51 & 8.0675 \\
Disclosure 256 & 4 & 53.11 & 52.02 & 0.4633 \\
\bottomrule
\end{tabular}
\vspace{3pt}\par\begin{minipage}{\linewidth}\footnotesize Same frozen distributions as the holdout comparison. Deviation is the largest joint-probability-weighted unilateral gain; training gains are zero within solver tolerance. Support size counts joint actions, not equilibria.\end{minipage}
\end{table}

\begin{figure}[!htbp]
\centering
\includegraphics[width=\linewidth]{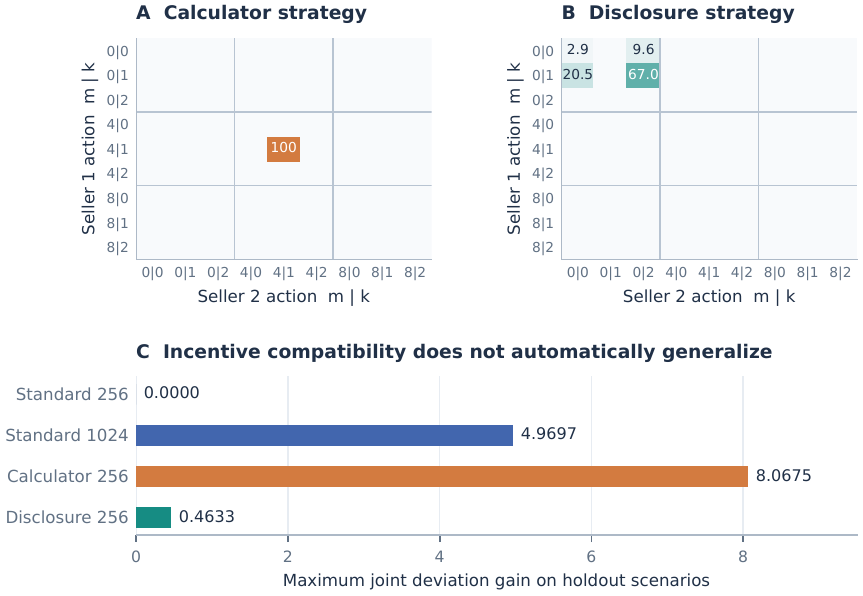}
\caption{\textbf{Training strategy support and holdout incentive violations.} A–B: probabilities (\%) of joint actions in selected consumer-worst training CE; m is monthly markup and k is complexity. Blank cells have zero mass. C: joint-probability-weighted unilateral gains on 16 holdout scenarios. Zero gain for standard 256 is compatible with low consumer surplus; it is not a quality score.}
\label{fig:4}
\end{figure}

The disclosure training distribution has four joint actions in its
support, all at zero markup. Its largest support probability is
approximately 66.97\%; the remaining probabilities are 20.48\%, 9.61\%,
and 2.94\%. The calculator distribution places all mass on the joint
action with markup 4 and complexity 1 for both sellers. The sparse
support is a property of the selected finite-game optimization result.
It should not be read as an estimate of how frequently actual firms
would adopt these precise formats.

Standard reasoning at budget 256 illustrates why incentive stability is
not a welfare ranking. Its selected distribution has zero holdout joint
deviation gain at numerical precision, yet holdout net consumer surplus
is -36.87. A buyer that makes sufficiently costly mistakes can support
an incentive-consistent but undesirable outcome. The comparison
therefore needs both the surplus measure and the deviation diagnostic
rather than a single composite claim that a strategy has generalized.

\hypertarget{markups-and-the-distribution-of-surplus}{%
\subsection{Markups and the distribution of
surplus}\label{markups-and-the-distribution-of-surplus}}

The calculator training distribution puts all probability on both
sellers selecting markup 4 and complexity 1. Every action in the support
of the disclosure training distribution has markup zero, with mixing
over a small set of representations. The full 48-scenario estimates
retain the difference between markup 4 and markup zero. Because zero
lies at the grid's lower boundary, this observation is conditional on
the available price actions.

Figure 5 evaluates the two buyer pipelines under the same fixed seller
distributions as a post hoc diagnostic. Under the calculator
distribution, disclosure earns 21.03 on the holdout scenarios and the
calculator earns 26.51. Under the disclosure distribution, they earn
52.02 and 49.83, respectively. Here \(W^h(q)\) averages over the 16
holdout scenarios, and \(q^C\) and \(q^D\) denote the distributions
selected on training scenarios. The overall difference can be decomposed
as

\begin{equation}
\label{eq:6}
\begin{aligned}W^D(q^D)-W^C(q^C)={}&\bigl[W^D(q^C)-W^C(q^C)\bigr]\\&+\bigl[W^D(q^D)-W^D(q^C)\bigr].\end{aligned}
\end{equation}

Along this path, \(25.51=-5.48+30.99\). Changing the seller distribution
first instead gives components 23.33 and 2.19. These terms depend on the
order of the decomposition. They are accounting comparisons of estimated
payoff matrices, not independently identified causal mediators; a
cross-evaluated distribution need not be an equilibrium for the buyer
pipeline under which it is evaluated.

\begin{figure}[!htbp]
\centering
\includegraphics[width=\linewidth]{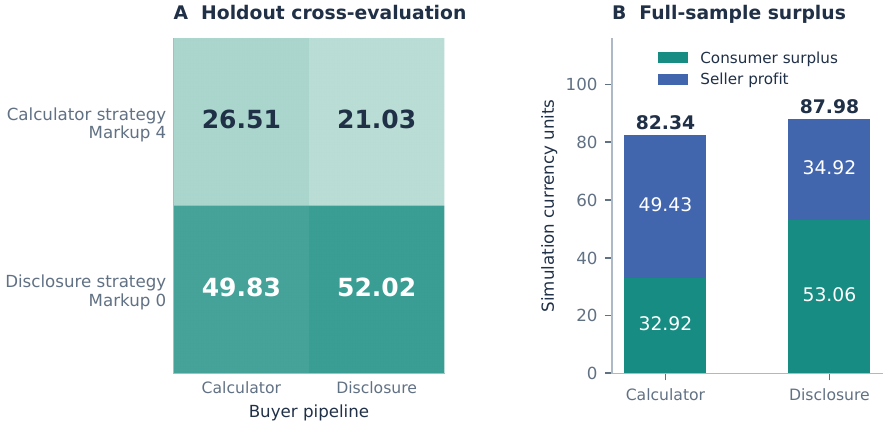}
\caption{\textbf{Seller strategies and the distribution of economic surplus.} A: post hoc holdout cross-evaluation; rows fix training distributions and columns change the buyer. Off-diagonal cells need not be equilibria. B: each 48-scenario consumer-worst CE; bar height equals net total surplus. The two panels use different estimation samples.}
\label{fig:5}
\end{figure}

In the full sample, the selected disclosure CE has 20.14 higher net
consumer surplus than the selected calculator CE, 14.51 lower aggregate
net seller profit, and 5.64 higher net total surplus. The
consumer-surplus difference combines changes in distribution,
allocation, and costs. The full 20.14 cannot be interpreted as a
resource saving. Welfare analysis must distinguish the incidence of
payments from changes in economic surplus.

Allowing negative markups, finer price steps, or additional
representations could alter best responses and equilibrium support. We
have no corresponding confirmation observations. The figure documents a
mechanism within the current finite action space, rather than a general
law of continuous-price competition.

\hypertarget{the-local-profitability-of-price-complexity}{%
\subsection{The local profitability of price
complexity}\label{the-local-profitability-of-price-complexity}}

The strategic interpretation can be examined without attributing every
change to equilibrium selection. Figure 6 takes a post hoc slice through
the measured payoff matrices. We hold the rival at the calculator
training action, markup 4 and complexity 1, and vary a seller's own
markup and complexity over all nine observed actions. For presentation,
each plotted value averages the two seller-specific net-profit matrices
over the full 48-scenario panel. The rival's strategy and the evaluation
sample are the same in both panels.

\begin{figure}[!htbp]
\centering
\includegraphics[width=\linewidth]{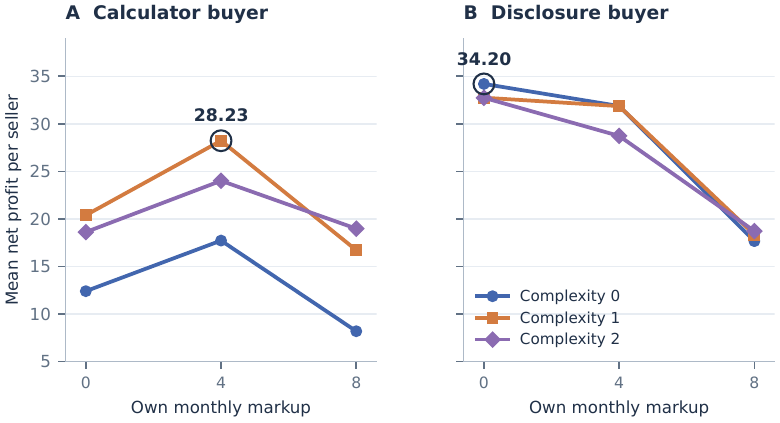}
\caption{\textbf{Residual errors change the profitability of price and format choices.} Post hoc slice of the 48-scenario payoff matrices: the rival is held at markup 4 and complexity 1. Values average the two seller-specific profit matrices, with $\kappa$ = 0.05. Circles outline each panel’s largest displayed value. This diagnostic does not solve a new equilibrium or identify a causal mediation effect.}
\label{fig:6}
\end{figure}

Under the calculator buyer, the largest mean profit in this slice is
28.23 at own markup 4 and complexity 1. At the same markup, direct
presentation yields 17.73 and complexity 2 yields 24.02. The ordering is
not monotone in our complexity index. Complexity 2 introduces additional
conversion operations, but it also changes the exact language of the
offer. A larger index therefore describes the designed transformation,
not a universal scalar measure of cognitive burden for the model.

Under disclosure, the largest displayed mean profit is 34.20 at zero
markup and complexity 0. With that representation held fixed, moving
from markup 4 to zero increases mean profit from 31.87 to 34.20 against
the specified rival. Under the calculator and complexity 1, the same
price move instead lowers mean profit from 28.23 to 20.40. These
opposite local payoff differences show how residual buyer errors can
coincide with different pricing incentives in this observed slice. They
are suggestive of a mechanism rather than an estimate of demand
elasticity or a causal channel.

The slice is deliberately limited. Averaging the sellers' profit
matrices is a descriptive display; each seller's own matrix, not the
pooled display, defines its equilibrium constraints. Moreover, the rival
is fixed at one action. Another opponent distribution can change the
ranking of own actions. The figure consequently supports a mechanism
interpretation within observed contracts and provides no independent
causal estimate of a general effect of complexity on profits.

\hypertarget{a-calculator-can-execute-the-wrong-contract-interpretation}{%
\subsection{A calculator can execute the wrong contract
interpretation}\label{a-calculator-can-execute-the-wrong-contract-interpretation}}

The calculator condition contains 1,075 invalid formats or expressions,
or 13.82\% of its observations. The prespecified outside-option fallback
happens to be optimal in 277 of these cases. A further 446 records are
valid but select the wrong option. Valid and optimal choices make up
80.44\% of all observations; accuracy conditional on validity is
93.34\%. The 84.00\% headline accuracy includes correct outside-option
fallbacks and should not be relabeled successful tool execution.

\begin{table}[!htbp]
\caption{Calculator output validity and economic correctness.}
\label{tab:6}
\centering\small
\setlength{\tabcolsep}{3.2pt}
\renewcommand{\arraystretch}{1.12}
\begin{tabular}{@{}>{\raggedright\arraybackslash}p{120.08pt}>{\centering\arraybackslash}p{55.75pt}>{\centering\arraybackslash}p{68.62pt}>{\raggedright\arraybackslash}p{184.42pt}@{}}
\toprule
\textbf{Outcome} & \textbf{Records} & \textbf{Share (\%)} & \textbf{Interpretation} \\ \midrule
Valid and optimal & 6,255 & 80.44 & Valid expressions, optimal choice \\
Valid but wrong & 446 & 5.74 & Valid expression, incorrect economic choice \\
Invalid but optimal & 277 & 3.56 & Outside-option fallback is optimal \\
Invalid and wrong & 798 & 10.26 & Outside-option fallback loses surplus \\
\bottomrule
\end{tabular}
\vspace{3pt}\par\begin{minipage}{\linewidth}\footnotesize 7,776 calculator decisions. Accuracy is (6,255 + 277) / 7,776 = 84.00\%; validity is (6,255 + 446) / 7,776 = 86.18\%. Percentages may differ slightly from 100 after rounding.\end{minipage}
\end{table}

Figure 7 distinguishes formal validity from economic correctness across
all four core pipelines. Standard and disclosure outputs are valid under
their answer parser in every recorded core decision. This does not mean
that their choices are correct. The calculator has a more demanding
output object: two expressions must be produced and accepted before the
deterministic comparison can proceed. Its lower formal-validity rate and
higher overall choice accuracy therefore describe different properties
of the complete procedure.

An alternative system might retry invalid expressions, escalate to a
human, or select a fallback model. Such a system would have a different
failure rate, cost profile, and demand response. The present experiment
measures the prespecified outside-option fallback. Correcting failed
calls after observing them would require accounting for the extra
computation and potentially solving the seller game again. It would not
be a harmless cleaning step applied to the existing outcomes.

One holdout example makes the remaining mechanism concrete. In scenario
10047, with both sellers at markup 4 and complexity 1, true total
payments are 166.8 and 288.0 and service values are 288 and 348. The
model supplies incomplete expressions that evaluate to 129 and 162. The
deterministic comparison then selects a seller whose true surplus is 60
rather than the alternative's 121.2. The same economic mistake occurs in
both positions, which remain two observations within one scenario.
Correct arithmetic execution does not ensure a correct semantic mapping
from contractual rules to arithmetic expressions.

\begin{figure}[!htbp]
\centering
\includegraphics[width=\linewidth]{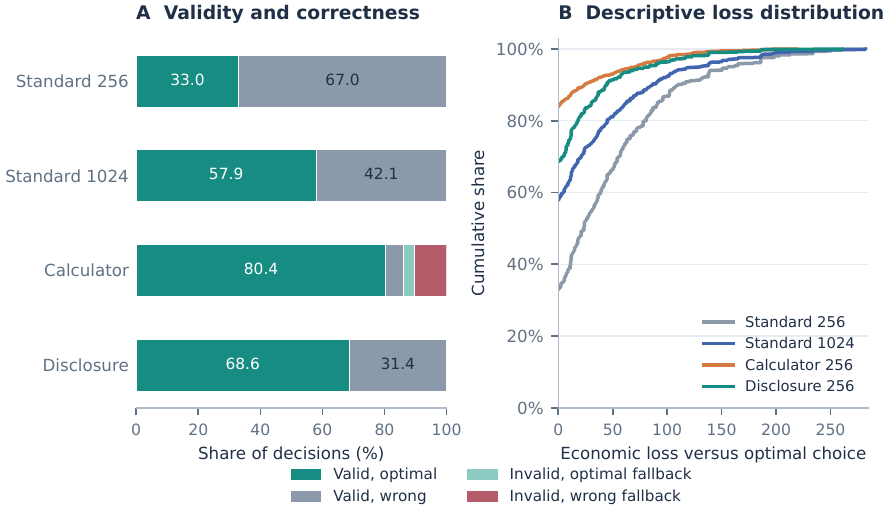}
\caption{\textbf{Reliable formatting and economic accuracy are distinct.} Post hoc descriptions of 7,776 decisions per core condition. A separates valid answers from correct fallback choices. B shows the empirical distribution of gross decision loss; curve colors follow the condition palette throughout the paper. Within-scenario observations are dependent, and no inferential bands are claimed.}
\label{fig:7}
\end{figure}

\hypertarget{loss-magnitudes-and-adverse-choices}{%
\subsection{Loss magnitudes and adverse
choices}\label{loss-magnitudes-and-adverse-choices}}

The calculator's static advantage is not confined to the zero-loss
event. Table 7 and Figure 7 report the severity of missed opportunities.
The average best attainable gross consumer surplus is 40.12 under the
uniformly weighted core menu; this benchmark is common to all four
conditions because they face the same underlying contracts and values.
Subtracting realized gross surplus gives mean losses of 42.51, 24.82,
8.22, and 13.23 for standard 256, standard 1024, calculator, and
disclosure, respectively. A mean loss above the benchmark is possible
because an incorrect buyer can accept a negative-surplus contract.

\begin{table}[!htbp]
\caption{Descriptive loss severity and adverse choices in the static menu.}
\label{tab:7}
\centering\small
\setlength{\tabcolsep}{3.2pt}
\renewcommand{\arraystretch}{1.12}
\begin{tabular}{@{}>{\raggedright\arraybackslash}p{104.02pt}>{\centering\arraybackslash}p{54.09pt}>{\centering\arraybackslash}p{66.57pt}>{\centering\arraybackslash}p{66.57pt}>{\centering\arraybackslash}p{66.57pt}>{\centering\arraybackslash}p{58.25pt}@{}}
\toprule
\textbf{Condition} & \textbf{Mean loss} & \textbf{95th pct. loss} & \textbf{99th pct. loss} & \textbf{Negative CS (\%)} & \textbf{No trade (\%)} \\ \midrule
Standard 256 & 42.51 & 156.00 & 234.00 & 40.57 & 25.27 \\
Standard 1024 & 24.82 & 129.22 & 199.80 & 20.61 & 31.30 \\
Calculator 256 & 8.22 & 70.20 & 126.00 & 3.56 & 35.44 \\
Disclosure 256 & 13.23 & 83.40 & 138.00 & 24.81 & 10.17 \\
\bottomrule
\end{tabular}
\vspace{3pt}\par\begin{minipage}{\linewidth}\footnotesize Post hoc descriptive summaries of the same 48 scenarios. Loss is best attainable gross consumer surplus minus realized gross surplus. Percentiles describe the record distribution; they are not confidence intervals or population value at risk.\end{minipage}
\end{table}

For a nonnegative decision loss \(L\) and optimal-choice accuracy
\(A<1\), the relationship between error frequency and error severity is

\begin{equation}
\label{eq:9}
\mathbb E[L]=(1-A)\,\mathbb E[L\mid L>0],\qquad A=\Pr(L=0).
\end{equation}

Two interventions with the same accuracy can differ in conditional loss
severity, and an accuracy improvement can coexist with larger losses on
the remaining errors. This identity motivates looking beyond a binary
score. It does not, by itself, explain the strategic reversal here: both
the calculator's mean loss and its reported upper-tail losses are lower
than disclosure's under the equal-profile menu.

The calculator's 95th and 99th loss percentiles are 70.20 and 126.00,
compared with 83.40 and 138.00 for disclosure. Negative gross surplus
occurs in 3.56\% of calculator decisions and 24.81\% of disclosure
decisions. The calculator also chooses no trade more frequently, 35.44\%
versus 10.17\%. These joint facts describe different error profiles.
They do not establish that every calculator abstention is desirable,
because foregoing a positive-surplus offer also creates decision loss.

The loss curves are descriptive empirical distributions over the
complete experimental menu. Repeated profiles within one scenario remain
dependent. We therefore do not present their upper percentiles as
population tail-risk estimates, and we do not call them financial value
at risk. Their role is to show that the strategic result cannot be
explained by pretending that disclosure dominates the calculator in
every dimension of fixed-menu decision quality. The reversal occurs
after seller-incentive constraints change how the offer space is
weighted.

\hypertarget{reading-the-static-and-strategic-ranking-together}{%
\subsection{Reading the static and strategic ranking
together}\label{reading-the-static-and-strategic-ranking-together}}

The fixed-menu result is a substantive finding in its own right. The
calculator pipeline not only has a higher probability of selecting an
optimal offer; it also has a lower mean decision loss and lower reported
upper-tail losses on the same equally weighted menu. This is more
informative than a narrow win on a binary accuracy score. It indicates
that, for the particular distribution of profiles chosen by the
evaluator, the calculator's mistakes are less costly on average and in
the descriptive tails. The result should not be weakened merely because
the strategic comparison orders the pipelines differently.

The strategic result answers a different question because the two
estimated games assign different economic significance to action
profiles. The consumer-worst CE is chosen separately from each policy's
set of incentive-consistent distributions. The calculator and disclosure
values therefore use different joint seller distributions as well as
different buyer responses. The 20.14-unit difference cannot be
attributed to a change in the buyer alone with seller actions held
constant, and it is not the average treatment effect of switching a
fixed market from one pipeline to the other.

Table 11 makes the change in weighting explicit. Relative to equal
action weights, the selected CE value is 1.48 units higher for the
calculator and 26.58 units higher for disclosure. The difference between
those two changes is 25.10 units, which accounts algebraically for
moving the disclosure-minus-calculator comparison from -4.96 to 20.14.
The equality holds by construction. It is informative about the
arithmetic of the reversal, but the two policy-specific reweighting
terms are not experimentally identified mediators and should not be
interpreted as effects of independently manipulating seller behavior.

A reader might ask whether a single shared seller distribution could
reverse the static ranking. The uniform menu is one common distribution,
and the post hoc holdout cross-evaluation uses two additional training
distributions. Under the calculator distribution the calculator has the
higher holdout surplus; under the disclosure distribution disclosure has
the higher holdout surplus. Those comparisons demonstrate heterogeneity
across the selected weights in this sample. They do not establish a
complete ranking under every feasible common distribution, and neither
off-diagonal evaluation is asserted to satisfy the equilibrium
conditions for the alternative buyer pipeline.

The careful interpretation is therefore conditional and comparative. The
observed response matrices admit a ranking reversal between the
equal-weight benchmark and the respective consumer-worst empirical CE
evaluations. The result supports the methodological argument that
evaluator-chosen weights and incentive-consistent weights can lead to
different welfare conclusions. It does not show that one policy
universally dominates across markets, that sellers actually coordinate
on the selected strategies, or that the same reversal survives every
perturbation of the scenario distribution.

\hypertarget{consumer-gains-seller-profits-and-efficiency}{%
\subsection{Consumer gains, seller profits, and
efficiency}\label{consumer-gains-seller-profits-and-efficiency}}

The welfare breakdown is economically important because a
consumer-surplus advantage can arise through both transfers and resource
use. For the selected full-sample CE distributions, disclosure's
consumer surplus exceeds the calculator's by 20.14, while the sum of
seller profits is lower by 14.51. Their sum is 5.64 higher after
rounding. The consumer gain is thus larger than the corresponding change
in the model's total surplus. Presenting only the consumer number would
conceal the distributional incidence; presenting only the total would
conceal who receives the gains.

The transfer identity does not imply that all price changes are socially
irrelevant. Holding the allocation fixed, a payment from the buyer to a
seller cancels when the two parties' payoffs are added. But a change in
price can alter which service is selected, whether trade occurs, and
which production cost is incurred. A change in format can alter
representation cost, and a different pipeline can alter token use. These
margins enter the model's total surplus and can change when the selected
distribution changes.

The calculations also do not establish that a 5.64-unit increase is a
complete social gain. There are only two sellers and one modeled buyer,
values are supplied by the generator, and both sides have quasilinear
payoffs. We do not include platform profits, externalities, quality
adjustments, entry costs, public finance, or the resources required to
generate verified totals. A regulator or platform would need to decide
how those missing components enter the objective and who bears the cost
before using a comparable calculation for policy.

The selected zero-markup disclosure actions are economically suggestive
but not a prediction that disclosure drives all real markups to zero.
Zero is the lower boundary of the available grid, and markup is added to
an underlying base charge rather than being the seller's entire
price-cost margin. A zero markup can coexist with positive profit, as
Table 3 and Figure 5 illustrate. Allowing negative markups, finer
increments, alternative products, or entry would change the set of
possible deviations and could change the computed equilibrium.

The holdout cross-evaluation is best interpreted as a diagnostic of how
buyer and offer distributions jointly enter outcomes. Starting from the
calculator distribution, switching only the pipeline changes average
holdout surplus by -5.48, and subsequently switching to the disclosure
distribution adds 30.99. Reversing the order produces different
components. The path dependence is a property of a two-factor accounting
decomposition with interaction; it is not evidence that one component is
the unique causal mechanism. The exercise nevertheless makes clear why
an evaluation based solely on a fixed menu need not predict a comparison
under policy-specific seller distributions.

\hypertarget{error-severity-nontrade-and-protocol-validity}{%
\subsection{Error severity, nontrade, and protocol
validity}\label{error-severity-nontrade-and-protocol-validity}}

The binary accuracy measure gives the same score to every nonoptimal
choice, whether its loss is small or large. The reported loss
distribution instead measures the distance between achieved gross
surplus and the best available gross surplus on the same menu. Because
the outside option is included in the benchmark, the loss is
nonnegative. A high value may reflect buying a poor contract or failing
to take a profitable one. The percentiles summarize the observed records
and are not estimates of a population tail-risk measure.

No-trade rates need equally careful interpretation. The calculator
chooses or is assigned the outside option more often than disclosure in
the static menu, but avoiding trade is not automatically a consumer
failure. If both contracts have negative surplus, no trade is optimal.
If one contract has positive surplus, abstention sacrifices an
opportunity. An invalid output mapped to the outside option can fall
into either category. The validity table separates whether the required
output was accepted from whether the final economic decision was
optimal.

This separation is especially important for a tool-assisted agent. A
valid calculator expression can be incomplete, while an invalid
expression can lead to an economically optimal fallback. Calling the
overall 84.00\% figure the rate of successful tool execution would
conflate these events. The reported 80.44\% valid-and-optimal share,
5.74\% valid-but-wrong share, and two invalid categories make the
pipeline's failure modes more explicit. They are descriptive categories
of the same 7,776 calculator decisions, not additional observations.

A deployment-oriented extension would compare complete recovery
strategies. Re-asking for an expression, verifying a term against a
structured contract, invoking an independent parser, or escalating to a
person could reduce some failures, but each could increase cost or
introduce new errors. An economic evaluation would need to record every
attempt and its cost, include all failures and fallbacks, and rebuild
the seller-response surface for the revised procedure. Correcting only
the failed cases after inspecting the answer would not measure the
performance of a prespecified deployable system.

\hypertarget{robustness-and-reasoning-budgets}{%
\section{Robustness and Reasoning
Budgets}\label{robustness-and-reasoning-budgets}}

\hypertarget{local-stability-and-joint-resampling}{%
\subsection{Local stability and joint
resampling}\label{local-stability-and-joint-resampling}}

The disclosure-minus-calculator CE lower-bound difference is positive in
all 48 leave-one-scenario-out estimates, ranging from 17.20 to 21.36. It
is also positive at every combination of \(\kappa\in\{0,0.05,0.2\}\) and
\(\lambda\in\{0,0.001\}\) (Table 8). At \(\lambda=0\) and
\(\kappa=0.05\), the difference is 20.09, so the main point estimate is
not principally driven by the computational-cost deduction.

\begin{table}[!htbp]
\caption{Sensitivity to representation and computational costs.}
\label{tab:8}
\centering\small
\setlength{\tabcolsep}{3.2pt}
\renewcommand{\arraystretch}{1.12}
\begin{tabular}{@{}>{\raggedright\arraybackslash}p{42.25pt}>{\centering\arraybackslash}p{50.70pt}>{\centering\arraybackslash}p{114.07pt}>{\centering\arraybackslash}p{114.07pt}>{\centering\arraybackslash}p{101.39pt}@{}}
\toprule
\textbf{$\kappa$} & \textbf{$\lambda$} & \textbf{Calculator lower CS} & \textbf{Disclosure lower CS} & \textbf{Difference} \\ \midrule
0 & 0 & 33.35 & 54.22 & 20.87 \\
0 & 0.001 & 32.89 & 53.81 & 20.92 \\
0.05 & 0 & 33.38 & 53.47 & 20.09 \\
0.05 & 0.001 & 32.92 & 53.06 & 20.14 \\
0.2 & 0 & 35.19 & 53.50 & 18.32 \\
0.2 & 0.001 & 34.73 & 53.09 & 18.36 \\
\bottomrule
\end{tabular}
\vspace{3pt}\par\begin{minipage}{\linewidth}\footnotesize Each row re-solves the CE programs on the same 48 scenarios. Positive estimates across this grid do not establish significance for a population equilibrium.\end{minipage}
\end{table}

Jointly resampling scenarios and re-solving the games produces 2.5th and
97.5th percentiles of -13.32 and 33.77 for the same contrast. Removing
one scenario is a local perturbation; it cannot substitute for changes
in multiple entries of the estimated payoff matrices. Changes in
equilibrium support and sample composition allow local stability and
wider resampling uncertainty to coexist.

The primary holdout difference is positive in 13 scenarios and negative
in three. Post hoc removal of one holdout scenario gives means between
19.93 and 28.91, and an ordinary paired-\(t\) interval is {[}8.02,
43.01{]}. These checks do not increase the number of independent
observations. The full-sample CE comparisons against standard reasoning
at 1,024 are more stable: re-solved sensitivity intervals are {[}16.61,
50.30{]} for the calculator and {[}19.71, 69.03{]} for disclosure.

\begin{figure}[!htbp]
\centering
\includegraphics[width=\linewidth]{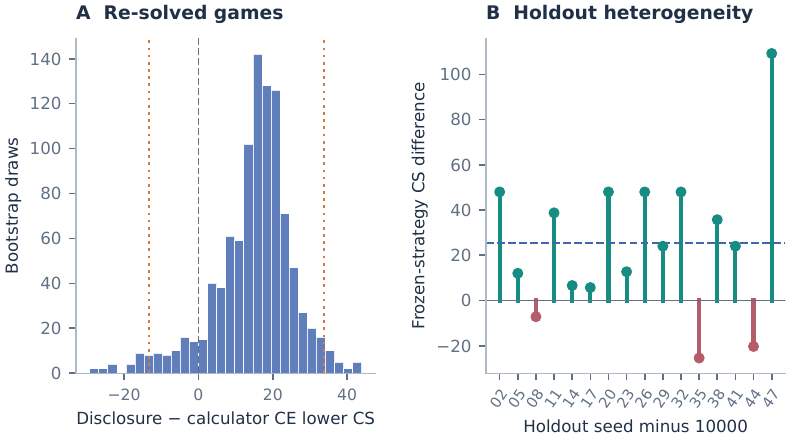}
\caption{\textbf{Joint perturbations and scenario-level heterogeneity.} A: disclosure-minus-calculator CE lower-bound differences; dotted lines mark the 2.5th and 97.5th percentiles. B: all 16 primary holdout differences; the dashed mean is 25.51. Thirteen are positive and three negative. Neither panel adds independent observations.}
\label{fig:8}
\end{figure}

\hypertarget{longer-reasoning-and-randomized-budgets}{%
\subsection{Longer reasoning and randomized
budgets}\label{longer-reasoning-and-randomized-budgets}}

All comparisons in Table 9 use the same prespecified 12 scenarios.
Standard reasoning at 8,192 achieves 86.93\% optimal choices and ends
naturally in 97.99\% of decisions, substantially reducing the truncation
seen at early low caps. Its empirical CE consumer-surplus range
nevertheless remains {[}-4.25, 23.81{]}. Natural completion and the
elimination of unfavorable strategic incentives are different
requirements.

\begin{table}[!htbp]
\caption{Matched-scenario comparison of reasoning budgets and interventions.}
\label{tab:9}
\centering\small
\setlength{\tabcolsep}{3.2pt}
\renewcommand{\arraystretch}{1.12}
\begin{tabular}{@{}>{\raggedright\arraybackslash}p{126.74pt}>{\centering\arraybackslash}p{76.05pt}>{\centering\arraybackslash}p{67.60pt}>{\centering\arraybackslash}p{84.49pt}>{\centering\arraybackslash}p{67.60pt}@{}}
\toprule
\textbf{Condition} & \textbf{Optimal (\%)} & \textbf{Gross CS} & \textbf{Token proxy} & \textbf{Net CS lower} \\ \midrule
Standard 256 & 29.99 & -21.58 & 398.50 & -61.69 \\
Standard 1024 & 50.15 & -2.04 & 1225.76 & -14.28 \\
Standard 8192 & 86.93 & 26.38 & 3801.16 & -4.25 \\
Calculator 256 & 82.41 & 24.76 & 463.90 & 18.14 \\
Disclosure 256 & 61.42 & 15.32 & 412.49 & 32.92 \\
Calibrated random & 42.93 & -10.68 & 1172.00 & -48.05 \\
Nominal random & 35.50 & -16.94 & 727.79 & -55.89 \\
\bottomrule
\end{tabular}
\vspace{3pt}\par\begin{minipage}{\linewidth}\footnotesize All rows use the same 12 scenarios. Random policies mix endpoint payoff matrices; they are not new observations. The token proxy is output tokens plus 0.1 times input tokens.\end{minipage}
\end{table}

\begin{figure}[!htbp]
\centering
\includegraphics[width=\linewidth]{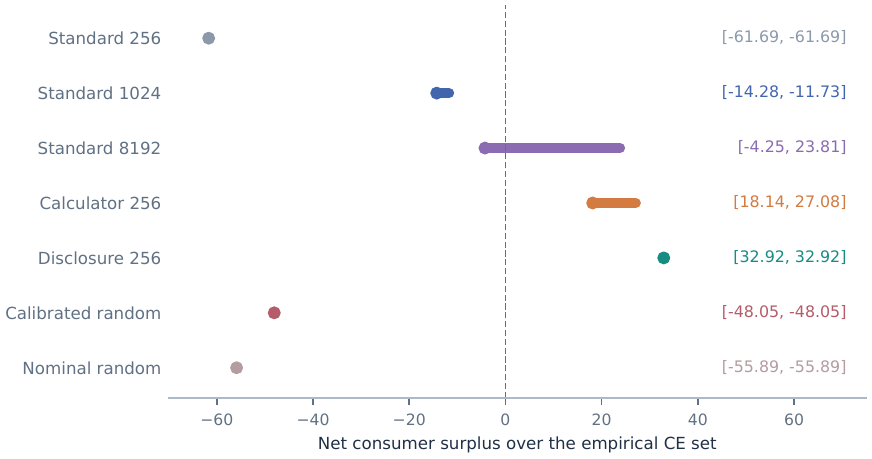}
\caption{\textbf{Equilibrium-set welfare ranges in the matched long-budget sample.} All seven policies use the same 12 scenarios. A segment connects the minimum and maximum within an empirical CE set; it is not a confidence interval. The point marks the lower bound. Random-budget matrices are mixtures of standard endpoint matrices.}
\label{fig:9}
\end{figure}

Development data fixed the high-budget probability at
\(\alpha=0.2273233\) when mixing caps 256 and 8,192 to match the token
proxy. A nominal-budget control uses
\(\alpha=(1024-256)/(8192-256)=3/31\). Randomization is realized after
seller actions and is not observable to sellers in advance.
Consequently, the relevant payoff matrix is the weighted average of the
two endpoint matrices. We do not treat the derived mixture as a newly
generated set of model decisions.

Relative to standard reasoning at 1,024, the calibrated random-budget
policy lowers the 12-scenario CE consumer-surplus lower bound by 33.77.
Its re-solved sensitivity interval is {[}-40.64, -0.60{]}, and all 12
leave-one-scenario-out differences are negative. Under the 8/4
training/holdout split, its frozen-strategy difference is -30.23, with
conditional interval {[}-46.61, -17.46{]}. Four holdout scenarios
support only a narrow conclusion.

Average static token-proxy costs are 1,172.00 for the calibrated mixture
and 1,225.76 for standard 1,024. Their amortized execution times are
1.325 and 0.820 seconds. Token matching on development scenarios is
neither exact matching on confirmation scenarios nor matching of GPU
time. These times divide batch runtime across decisions; they are not
isolated end-user request latencies.

Within the same 12-scenario panel, the four-scenario
disclosure-minus-calculator holdout difference is -1.22, with interval
{[}-38.24, 44.45{]}. The fitted strategies and sample differ from those
of the core comparison. This result is neither an independent
replication nor a clean contradiction, but it further limits a universal
dominance claim. The original four-scenario pilot produced a
random-budget advantage of approximately 47.04; subsequent 8B
calibration and confirmation-subset differences are -31.89 and -33.77.
Budgets, protocols, and scenarios changed across stages, so these
estimates are not one causal time series. Their directional change
remains part of the research record.

\hypertarget{resource-use-and-the-limits-of-a-single-frontier}{%
\subsection{Resource use and the limits of a single
frontier}\label{resource-use-and-the-limits-of-a-single-frontier}}

Figure 10 organizes the matched 12-scenario evidence by resources. The
left panel relates mean token use to static optimal-choice accuracy. The
right panel relates amortized execution time to the empirical CE lower
bound on net consumer surplus. These are separate resource measures and
separate performance objectives. A point with greater accuracy need not
have the higher strategic surplus, and a mixture matched on tokens need
not be matched on seconds.

\begin{figure}[!htbp]
\centering
\includegraphics[width=\linewidth]{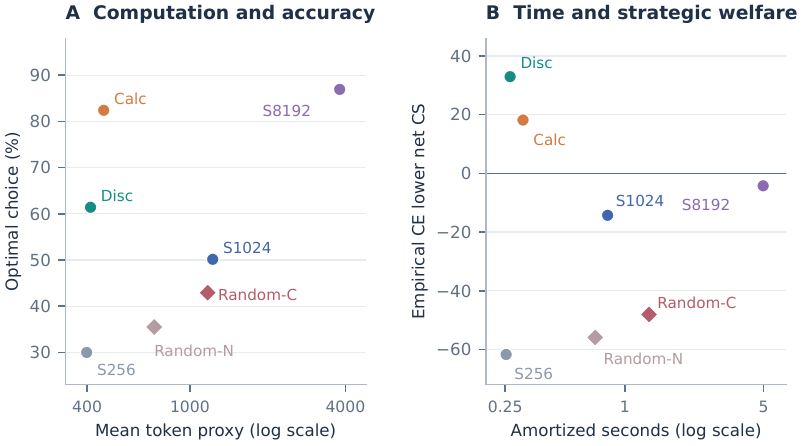}
\caption{\textbf{Compute, accuracy, and strategic welfare need separate measurement.} Descriptive resource–performance maps on the same 12 scenarios. S denotes standard reasoning; Calc and Disc denote calculator and disclosure. Random-C uses the calibrated mixture and Random-N the nominal mixture. Time is batch runtime allocated per decision, not isolated request latency. Points are evaluated policies, not an estimated production frontier.}
\label{fig:10}
\end{figure}

Standard reasoning at 8,192 uses a token proxy of 3,801.16 and
approximately 4.970 amortized seconds per decision. The calculator uses
463.90 and 0.307 seconds on the same scenarios. The longer standard
procedure delivers higher static accuracy, 86.93\% versus 82.41\%, while
its lower CE surplus is -4.25 rather than 18.14. This is not a universal
comparison between large and small reasoning budgets: only the endpoint
prompts, the observed model, and the selected scenarios are covered. It
is a comparison of the measured computational procedures in the
specified market.

Disclosure uses 412.49 token-proxy units and 0.265 amortized seconds in
this subset. Its favorable strategic point estimate does not imply that
producing verified totals is costless. Those upstream calculations are
supplied by the generator and omitted from the plotted execution time. A
deployment comparison would need an additional cost category for
acquiring, verifying, and updating those inputs. Treating the plotted
buyer runtime as the entire cost of disclosure would bias a procurement
decision.

We avoid fitting a smooth production frontier through the seven points.
Several policies differ in information inputs and decision rules, and
the random-budget points are convex combinations of endpoint payoff
matrices rather than independent measurements of a learned scheduler. A
fitted curve could conceal those design differences. The scatter plot
instead exposes which combinations of cost and outcome were actually
evaluated and which prospective systems remain unobserved.

\hypertarget{what-the-negative-budget-result-rules-out}{%
\subsection{What the negative budget result rules
out}\label{what-the-negative-budget-result-rules-out}}

The failed random-budget intervention narrows the economic claim.
Randomizing computational effort does not automatically protect
consumers simply because sellers cannot observe the realized budget in
advance. Sellers respond to the expected payoff matrix generated by the
mixture. If the low-budget component leaves sufficiently profitable
mistakes, occasional high-budget decisions need not remove the
incentives supporting an unfavorable outcome. The measured mixture
demonstrates this possibility in the present sample; it does not
identify the optimal mixing probability.

Selecting the best probability after examining confirmation welfare
would create a new, data-dependent policy. Its performance would require
a new holdout set or an appropriate nested evaluation. We therefore keep
the development-calibrated probability fixed and retain its negative
result. The same principle applies to choosing a more favorable
complexity grid or deleting a difficult scenario after seeing the
equilibrium estimate. Such changes are reasonable research directions
but cannot be relabeled confirmations of the original plan.

\hypertarget{what-the-uncertainty-exercises-do-and-do-not-establish}{%
\subsection{What the uncertainty exercises do and do not
establish}\label{what-the-uncertainty-exercises-do-and-do-not-establish}}

The conditional holdout interval treats the two training-selected
distributions as fixed and resamples the paired scenario-level surplus
differences on the holdout set. This construction respects the common
scenario primitives across buyer conditions. Its interpretation is
conditional on the distributions that happened to be fitted in training
and on the synthetic scenario-generating process. It does not include
the variability that would arise if a different training sample produced
different seller payoff matrices and a different selected equilibrium.

The re-solved exercise changes more than the holdout resampling step.
Each resampled game is built from a new collection of scenarios and the
CE programs are solved again. The optimizer may move to a different
extreme point or its support may change. The reported 2.5th and 97.5th
percentiles, -13.32 and 33.77 for the full-sample
disclosure-minus-calculator comparison, include zero. We use this range
to show sensitivity of the computed difference, not as a guaranteed
confidence interval for a population equilibrium. The non-smooth
optimization and the limited sample make a formal coverage claim
inappropriate without further theory.

The leave-one-scenario-out results provide another, narrower diagnostic.
Every reported deletion gives a positive full-sample lower-bound
difference, but this does not contradict a resampling range that
includes negative values. Removing one observation perturbs the data
locally. A bootstrap draw can repeat several scenarios and omit others,
changing multiple payoff entries simultaneously. Robustness to any
single deletion is therefore not the same as robustness to a broader
change in the empirical scenario distribution.

Model-generation uncertainty is distinct again. For each scenario,
action profile, condition, and offer position, the confirmation design
records one response under its generation protocol. Scenario resampling
groups those recorded responses with their scenario, but it does not
reveal the full variation that repeated independent generations would
create for a fixed prompt. In particular, a repeated run might alter an
action's estimated profit enough to affect an incentive constraint or
selected support. A future design should vary both scenario draws and
generation seeds and specify how these sources of uncertainty enter the
equilibrium calculation.

The positive holdout result and the wider re-solved range can therefore
coexist without one invalidating the other. The former is favorable
evidence for the fitted distributions on the specified holdout
scenarios; the latter cautions that a market-equilibrium ranking is less
stable when the fitted game is perturbed and reoptimized. The deviation
diagnostic adds a third question: whether a frozen distribution still
satisfies seller incentives under holdout-average payoffs. It need not
do so even when its consumer-surplus comparison transfers favorably.

\hypertarget{interpreting-the-randomized-budget-comparison}{%
\subsection{Interpreting the randomized-budget
comparison}\label{interpreting-the-randomized-budget-comparison}}

Randomizing a reasoning budget is attractive because the buyer can
occasionally use a more capable, costly procedure without paying its
cost on every decision. In the measured comparison, the random policy is
represented by a mixture of the short- and long-budget payoff matrices.
The selected high-budget probability was fixed using development
information, and the mixture is evaluated on the later matched
scenarios. The high-budget draw is modeled as occurring after seller
actions and being unobserved by sellers in advance, so seller incentives
respond to the expected payoff matrix of the mixture.

This construction is not the same as asking a language model to choose
its own budget based on a contract. A learned adaptive policy could
allocate more computation to difficult or economically consequential
offers; it could also be manipulated if sellers can influence the signal
used to allocate resources. Such a policy would need a prespecified
selection rule, a record of all computational costs, and new evaluations
for every relevant seller action. The present mixture does not establish
the value of adaptive allocation or an optimal probability of using the
long budget.

On the 12-scenario matched sample, the calibrated random policy has a
lower empirical CE consumer-surplus bound than standard reasoning at
1,024 tokens. Its re-solved sensitivity range for that contrast remains
negative in the reported calculation. This is a result for the measured
mixture and specified subset, not a claim that every random-budget
design reduces welfare. The subset overlaps the main scenarios, has only
four held-out scenarios in its split, and inherits the same uncertainty
about game estimation and equilibrium selection.

The economic explanation is plausible without being separately
identified. If the low-budget component leaves a profitable demand
response at some actions, weighting it with a high-budget component need
not remove that incentive. The effect depends on the magnitude and
location of the remaining errors and on the sellers' margins, not merely
on expected token consumption. A cost-matched average can therefore have
a different strategic value from a deterministic intermediate budget
even when their total token proxies are similar.

This finding also explains why more compute is not an automatic
substitute for a well-designed decision interface. The 8,192-token
standard procedure improves fixed-menu accuracy and reduces truncation
in the matched panel, but its empirical CE welfare range remains broad
and includes a negative lower bound. Completion, exact-choice accuracy,
expected loss, and seller incentives are separate outcomes. A reliable
evaluation should measure each rather than infer them from the reasoning
cap alone.

\hypertarget{economic-interpretation-and-implications}{%
\section{Economic Interpretation and
Implications}\label{economic-interpretation-and-implications}}

\hypertarget{buyer-technology-and-endogenous-offer-composition}{%
\subsection{Buyer technology and endogenous offer
composition}\label{buyer-technology-and-endogenous-offer-composition}}

In a pricing game, the economically relevant object is the demand
response to each feasible combination of price and presentation. Buyer
accuracy summarizes one aspect of that response under a particular
evaluation distribution. Seller profit depends on how often a given
action attracts the buyer and on the payment and cost attached to that
action. A small number of mistakes at profitable action profiles can
therefore matter disproportionately for a seller's incentives.
Conversely, many errors on profiles with little equilibrium weight may
have limited influence on a selected equilibrium outcome.

The experiment provides the ingredients for that argument: a complete
finite grid, a buyer choice for each profile and position, and seller
payoffs implied by the selected contract. The CE solution uses the
payoff matrices to select incentive-consistent joint distributions. The
observed change in weights is a model-based counterfactual, not a
measured process in which firms actually learn or change their prices.
That distinction is necessary for interpreting the result as industrial
organization rather than a field estimate of market adjustment.

The concept also clarifies the role of price complexity. In the
exact-comparison benchmark, representation changes neither demand nor
revenue at a fixed markup, so positive complexity is strictly dominated
when it is costly. With an imperfect AI buyer, presentation can change
demand even though the underlying payment is unchanged. Complexity is
then a potential strategic instrument. Our three formats permit a
controlled comparison, but their numbering does not define a universal
scale of model difficulty, and the observed payoff differences need not
generalize to other wording or sellers.

\hypertarget{benchmark-design-and-strategic-reweighting}{%
\subsection{Benchmark design and strategic
reweighting}\label{benchmark-design-and-strategic-reweighting}}

A fixed-menu benchmark is appropriate for measuring how often a pipeline
reaches an optimal decision on that menu. It is also valuable for
locating contract-specific failures. The strategic calculation answers a
different question: which outcomes remain feasible when seller
deviations are constrained by the measured buyer response? The two
answers use the same model calls but attach different weights to the
action space. Neither accuracy nor the uniform-menu surplus difference
determines the CE comparison by itself.

The reweighting identity and Table 11 make this distinction explicit.
For a proposed system, an evaluator can report uniform-menu performance,
a clearly specified strategic outcome, and the change between them. When
the strategic object is an equilibrium set, the equilibrium concept,
selection criterion, and range of feasible seller actions should be
reported. A single consumer-worst CE is informative as a conservative
model-based criterion, but it does not identify which equilibrium a
market would select.

The same discipline applies to generalization. Evaluating a frozen
distribution on new scenarios asks whether that particular distribution
transfers. Checking its deviation gains asks whether it remains
incentive-compatible under the new payoffs. Re-solving the game on
resampled scenarios examines sensitivity to changing payoffs and
equilibrium selection. These are complementary exercises, and their
uncertainty intervals should not be treated as interchangeable.

\hypertarget{incidence-efficiency-and-the-cost-of-information}{%
\subsection{Incidence, efficiency, and the cost of
information}\label{incidence-efficiency-and-the-cost-of-information}}

The distinction between private incidence and aggregate efficiency is
central to welfare economics. Contract payments are transfers between
the buyer and the selected seller in our accounting. The sum of consumer
surplus and seller profit removes those payments and retains the value
of the chosen service, its resource cost, computation cost, and the
costs of the sellers' representations. A lower payment can substantially
improve consumer surplus even when the change in total surplus is much
smaller.

For the selected 48-scenario equilibria, disclosure's 20.14-unit
consumer advantage is accompanied by 14.51 lower seller profit and 5.64
greater total surplus. These are model-implied differences, not a
complete welfare analysis of a real market. Product availability,
quality, entry, and supplier investment are fixed, and no distributional
weights are estimated. The calculation nevertheless prevents the
consumer gain from being presented as an equally large gain in aggregate
resources.

Information provision also uses resources. The verified totals are
supplied by the generator; the experiment charges the buyer's
token-based proxy but does not measure the upstream cost of generating,
checking, updating, or correcting totals. A deployment-level assessment
would need those costs as well as potential errors in the information
itself. With fixed verification cost \(F\) shared across \(n\)
decisions, \(F/n\) is one relevant additional cost per decision, but the
appropriate net effect also depends on market behavior and
implementation. The study's 25.51-unit holdout difference is not an
estimate of this deployment benefit.

\hypertarget{consumer-markets-and-a-bounded-finance-application}{%
\subsection{Consumer markets and a bounded finance
application}\label{consumer-markets-and-a-bounded-finance-application}}

The direct setting is a synthetic subscription market, which is relevant
to a range of recurring-fee contracts. Payment and deposit accounts can
also combine fixed charges, transaction fees, and conditions for
waivers. This similarity suggests a testable application to financial
services. It does not turn the present synthetic contracts into
observations of actual bank fees, borrowers, investors, or regulatory
outcomes.

An application to financial products would require independently
verified contract terms, economically relevant customer profiles, and a
clear objective for comparing products. For credit, repayment timing and
borrowing constraints matter; for investment or insurance, risk and
state-contingent outcomes matter. The current verified-total
intervention supplies only a deterministic payment statistic. It cannot
establish product suitability or substitute for those additional
components of the decision.

\hypertarget{implications-for-evaluation-and-further-research}{%
\subsection{Implications for evaluation and further
research}\label{implications-for-evaluation-and-further-research}}

An economic evaluation of an AI buyer should retain the raw contract,
extracted terms, intermediate calculations, final choice, and resulting
payoff. It should measure decision loss in addition to exact-choice
accuracy and preserve outside-option and invalid-response outcomes. If
an alternative design introduces retries, a fallback model, or human
review, its incremental cost and its effect on the demand response
should be measured rather than assumed.

Strategic evaluation should define the actions suppliers can plausibly
change and test whether results persist near action-grid boundaries.
Further studies can add repeated model samples, independently selected
scenarios, other model families, richer price grids, and alternative
solution concepts or observed adjustment processes. A common-interface
ablation would distinguish providing verified totals from changing the
final decision procedure. External product and pricing data would then
permit a more credible assessment of transfer to a specific market.
These extensions address distinct sources of uncertainty; none is
supplied by the present confirmation panel.

\hypertarget{buyer-adoption-and-the-market-for-decision-assistance}{%
\subsection{Buyer adoption and the market for decision
assistance}\label{buyer-adoption-and-the-market-for-decision-assistance}}

Our model compares markets in which a single specified buyer pipeline is
used consistently. This isolates how a pipeline maps a common action
grid into choices and seller profits. Actual adoption is likely to be
heterogeneous: some customers may use an agent, others may compare
offers manually, and others may accept a recommendation from a platform.
With heterogeneous buyers, a seller's expected demand would average the
responses of those groups, and the incentive to choose a particular
format would depend on their shares as well as on their margins.

The existing data cannot estimate that adoption mixture because they do
not observe human responses or a population of agent users. Combining
the observed AI payoff matrices with an assumed second population would
define a new counterfactual model, but its results would inherit the
assumed response and adoption shares. If sellers can identify which
buyer uses which system, they might offer different terms to the two
groups; if they cannot, the same action may have to attract both. Those
information and segmentation assumptions would have to be stated before
calculating the corresponding equilibrium.

There is also an upstream market for the assistance itself. A buyer may
pay a provider, a platform may supply the agent without a visible fee,
or a seller may influence how a shopping tool is financed. The present
computational-cost proxy is charged directly in the buyer's net surplus
and does not model a provider's price, fixed investment, advertising
revenue, or incentives to recommend one seller. Those issues are
economically important, but none is identified from the two-seller
product game studied here.

Competition among assistance providers could change which systems are
available and which accuracy-welfare tradeoffs consumers face. A
provider rewarded only for benchmark success might select a different
interface from one rewarded for realized consumer surplus, especially if
the benchmark's offer distribution differs from deployment. Conversely,
a provider may have limited ability to change sellers' contract terms.
An empirical study of that market would need both provider-side
incentives and evidence on how sellers respond to the deployed tool;
neither can be inferred from the model's selected CE alone.

These distinctions suggest a broader research agenda rather than an
additional result. One can first estimate response surfaces for several
buyer pipelines and human or non-agent benchmarks; then specify the
share of each buyer type, the information available to sellers, and
whether prices can be segmented. Only after these choices can the
corresponding mixed-population game be solved. Reporting the assumptions
separately would show which policy conclusions are driven by measured
behavior and which are driven by the adoption model.

\hypertarget{interpretable-information-and-the-institutional-cost-of-disclosure}{%
\subsection{Interpretable information and the institutional cost of
disclosure}\label{interpretable-information-and-the-institutional-cost-of-disclosure}}

The experiment supplies verified total payments as an input, but the
economic institution that supplies them remains outside the model. In a
recurring-fee market, computing a total may require a usage assumption,
an account horizon, a rule for introductory promotions, and an
interpretation of conditional credits. A single number can be useful
when those inputs are agreed upon, yet the relevant total can change
when a consumer's future behavior differs from the assumed profile. The
synthetic scenarios avoid this ambiguity by giving the model a known
horizon and usage level.

A practical disclosure design would have to specify whether a provider
reports a total for each customer's profile, a standardized illustrative
profile, or a range under different usage assumptions. These
alternatives answer different questions and incur different data and
verification costs. A standardized figure may improve comparability
while omitting customer-specific information; a personalized calculation
may require data access and ongoing updates. We do not compare these
designs or estimate their costs in the present experiment.

The method can nonetheless guide evaluation. First, measure how a
buyer's choices change when the information is supplied under a
specified profile and when it is absent. Second, record the costs and
error rates of producing that information. Third, examine the sellers'
feasible responses, including changes to the terms or representations
that the disclosure must summarize. Finally, state whose surplus is
affected and whether a change in payments is a transfer or a change in
resource use. Without these steps, a reported improvement in task
accuracy is not enough to establish the net benefit of a disclosure
institution.

The distinction is useful for financial applications but does not make
the current experiment a financial-market study. Payment accounts,
credit contracts, investment products, and insurance policies differ in
timing, risk, regulation, and suitability. A correct statement of a
deterministic payment may be only one input to a credit decision and may
not capture default risk, liquidity, or the distribution of outcomes.
Extending the design would require verified product terms and a clearly
specified customer objective, not simply renaming the synthetic
subscriptions as financial products.

Similarly, the observed disclosure comparison does not identify a
regulation. A rule that requires sellers to provide information may
change firms' compliance costs and their ability to select or update
contract terms. It could also require enforcement or allow strategic
omissions not represented in our verified input. Our model can make such
assumptions explicit in future work, but the current evidence does not
estimate a regulatory effect or establish that a particular policy
should be adopted.

\hypertarget{evaluating-a-buyer-as-a-market-institution}{%
\subsection{Evaluating a buyer as a market
institution}\label{evaluating-a-buyer-as-a-market-institution}}

An AI buyer can be assessed at several levels. At the first level, a
researcher checks whether the system returns an accepted response. At
the second, the researcher checks whether that response is economically
correct for a particular menu. At the third, the researcher assigns
losses and costs to the alternatives. At the fourth, the researcher asks
which menus remain incentive-consistent when counterparties can alter
their actions. Each level adds information; none can generally be
reconstructed from the level before it.

This hierarchy has a practical consequence for benchmark construction. A
benchmark that samples offers uniformly can reveal failures across a
controlled action grid and enable comparisons between systems. Its
aggregate score should be accompanied by action-level outcomes and
contract primitives so that alternative economically relevant weights
can be examined. A benchmark that uses only historically observed
contracts may better describe a current market distribution, but it can
omit profitable deviations that firms have not yet tried. Combining
observed distributions with a defensible set of counterfactual offers is
a research design problem, not a purely technical scoring choice.

The stability of the outcome should also be evaluated separately from
its level. A low deviation gain means that, under the specified
empirical game and diagnostic, there is little
joint-probability-weighted incentive to move away from a recommendation.
It does not imply high consumer surplus. A high consumer-surplus value
does not imply that sellers have no profitable deviations. In the
present holdout, the two statistics reveal different properties of the
training-selected distributions. A report that collapses them into one
score would lose that distinction.

Distributional effects introduce another dimension. An identical
total-surplus change can be distributed differently across buyers,
sellers, and providers. Even within a consumer population, users with
different valuations, usage profiles, or fallback options may be
affected differently. The current sample contains artificial scenarios
rather than demographic groups, so it cannot support claims about
distribution across income, age, or other protected characteristics.
Such claims require appropriate population data, a sampling design, and
an explicit welfare criterion.

The central economic message is not that accuracy should be discarded.
Accuracy is a useful diagnostic, and in this study the calculator's
strong fixed-menu performance is worth reporting. The message is that
accuracy is one input into a welfare assessment. The interpretation of
an AI shopping system depends on the losses its remaining errors create,
the offers sellers choose in response, the costs of producing and
checking information, and the division of gains across market
participants. Those components are conceptually distinct even when they
are calculated from a common set of model outputs.

\hypertarget{limitations}{%
\section{Limitations}\label{limitations}}

The main evidence covers one 8B model, one English-language synthetic
contract domain, and a finite grid of seller actions. The 14B
observations belong to development calibration rather than formal
cross-model confirmation. The disclosure strategy has zero markup on its
selected support, at the lower boundary of the grid; negative markups or
a finer grid could change the result. The 12 long-budget scenarios
overlap the core panel and do not constitute an independent replication.

The empirical game averages one stochastic generation for each action
profile and each of two positions. Scenario clustering preserves the
dependence created by shared contract primitives, but it does not
identify the full distribution of repeated outputs from a fixed prompt.
Shared seeds across conditions do not guarantee identical random paths.
Additional generations could change estimated payoff matrices and their
equilibrium sets.

CE describes incentive-consistent distributions of the estimated finite
game. We do not observe sellers, a recommendation mechanism, learning,
or equilibrium selection. The frozen training distributions admit
positive holdout joint deviation gains in the calculator, disclosure,
and standard-1024 conditions. The full-sample re-solved sensitivity
range spans zero and has no established coverage theorem for this
non-smooth optimization. The fixed-strategy holdout interval conditions
on training distributions and excludes their estimation and selection
uncertainty.

The calculator and verified-total conditions differ in prompts,
information, and decision procedures. Their comparison does not identify
the isolated effect of one tool, one disclosure field, or a regulatory
policy. The verified inputs are error-free by construction; their
production and verification costs are not measured. The token-cost and
representation-cost parameters are illustrative rather than calibrated
estimates of real resource costs.

Finally, the economic values and costs are investigator-specified and
utilities are quasilinear. There are no observed households, financial
products, endogenous product quality, entry, credit risk, or dynamic
contracts. The outcomes are simulation units, not estimates of currency
losses or welfare effects in a particular consumer or financial market.

\hypertarget{scope-of-the-strategic-and-welfare-claims}{%
\subsection{Scope of the strategic and welfare
claims}\label{scope-of-the-strategic-and-welfare-claims}}

The modeled seller action is selected before a scenario is realized. A
different ordering would permit different forms of conditioning and
might generate a different equilibrium set. In particular, the
experiment does not show how firms would set prices after observing a
customer's full valuation or usage, nor does it estimate the information
that firms possess in actual markets. Both the scope of a strategy and
the information available when it is chosen are part of the model rather
than empirical facts inferred from AI responses.

The correlated-equilibrium calculation imposes incentive constraints on
expected seller payoffs for the specified action grid. It does not model
how a recommendation is generated, whether firms can communicate, or
which feasible equilibrium is selected. The consumer-worst criterion
supplies a conservative value within the fitted set, but it is not a
forecast that the market will select the worst outcome. Alternative
selection criteria, Nash equilibria, or explicit learning models could
give different comparisons and would answer different economic
questions.

The data do not isolate the causes of individual model mistakes. In
particular, the complexity index bundles changes in presentation and
numerical conversion, and the buyer conditions bundle prompts, inputs,
and decision procedures. The fixed-opponent slices are descriptive
comparisons, not randomized mediation estimates. A stronger mechanism
design would vary one contractual feature at a time and test whether the
resulting demand response survives when the remaining components and
seller incentives are held fixed.

The present uncertainty analysis also cannot rule out chance patterns
induced by a small number of difficult scenarios or stochastic
generations. Scenario-level resampling, leave-one-out calculations, and
a held-out split address parts of this concern, but the most favorable
interval conditions on the selected training distributions and the
re-solved range spans zero. More model draws per prompt, independently
generated scenarios, and a repeated train-test design would help
separate noise in the demand surface from variation in equilibrium
selection. They would not, by themselves, validate the assumptions of
the market model.

Finally, the evidence is generated from one English-language model and a
synthetic contract generator. The known valuations make welfare
calculations transparent, but they are chosen by the investigator rather
than recovered from observed willingness to pay. The simulated costs and
token prices are illustrative. These limitations mean that the results
should be read as a controlled computational demonstration and empirical
evaluation framework, not as a calibrated estimate of the magnitude of
welfare gains in an actual economy.

\hypertarget{conclusion}{%
\section{Conclusion}\label{conclusion}}

This study asks how the economic value of an AI buyer changes when the
offers it evaluates are subject to strategic seller incentives. In the
synthetic subscription market, the calculator achieves higher fixed-menu
accuracy than verified total-payment disclosure: 84.00\% versus 68.57\%.
It also has lower average decision loss on the equal-profile menu. These
findings establish a clear benchmark advantage for the complete
calculator pipeline on the observed contracts.

The comparison changes when the same responses define seller payoff
matrices and each intervention is evaluated over its empirical CE set.
The disclosure condition has a consumer-surplus lower bound of 53.06,
compared with 32.92 under the calculator. At uniform action-profile
weights the disclosure-minus-calculator difference in net consumer
surplus is -4.96; at the respective consumer-worst CE distributions it
is +20.14. The difference between these comparisons, 25.10, is an
accounting consequence of the policy-specific weights placed on the
measured payoff surfaces. This is the paper's main economic result: the
distribution used to evaluate an AI buyer can matter as much as its
aggregate task score.

The holdout evidence answers a narrower but useful question. When
training distributions are fixed and evaluated on 16 new scenarios,
disclosure delivers 25.51 more net consumer surplus, with a 95\% paired
interval of {[}10.69, 42.08{]} conditional on those distributions. The
distributions are not exact equilibria under holdout payoffs, and the
full-sample sensitivity range from re-solving resampled games is
{[}-13.32, 33.77{]}. The positive conditional holdout comparison should
therefore be read alongside, not as a replacement for, the uncertainty
in the equilibrium comparison.

The payoff results offer an interpretation of the reversal. The selected
calculator training distribution places all mass on a profile in which
both sellers choose markup 4 and complexity 1. The selected disclosure
distribution places its mass on zero-markup profiles. In a post hoc
cross-evaluation, changing the buyer pipeline while holding the
calculator distribution fixed reduces holdout surplus by 5.48; changing
the seller distribution afterward adds 30.99. Reversing the order yields
different components. These numbers demonstrate the dependence on offer
weights and evaluation path; they do not establish an independently
identified causal mechanism or show that real sellers would choose those
prices.

The economic importance of the result also depends on who gains. In the
selected full-sample equilibria, the 20.14-unit consumer-surplus
advantage is accompanied by 14.51 lower seller profit and 5.64 higher
total surplus. Because payments are transfers in the model, an increase
in consumer surplus need not represent an equal increase in aggregate
economic resources. Allocation, computation, representation costs, and
the distribution of payments all matter. This accounting provides a
clearer basis for welfare discussion than a comparison of accuracy or
prices alone.

The idealized representation-invariance result defines an important
boundary. If a buyer always computes payments and maximizes surplus
correctly, costly representation complexity is strictly dominated at a
given markup. The tested disclosure buyer is not such an ideal agent: it
receives correct payments but must still compare them with values and
choose. The calculator can execute an expression correctly even when the
expression encodes the wrong contractual rule. Both observations locate
the economic issue in the full decision procedure and in the particular
errors sellers face.

The negative randomized-budget finding gives a second warning against a
single capability metric. On the matched 12-scenario subset, the
calibrated mixture of short and long reasoning budgets does not improve
the consumer-welfare lower bound relative to standard 1,024-token
reasoning. The sample is small and overlaps the core panel, and the
mixture is constructed from endpoint payoff matrices. Within those
limits, the result shows that occasional high-budget decisions need not
remove the incentives associated with lower-budget mistakes.

These conclusions remain conditional on a single model, artificial
values and costs, one response per profile and position, and a
restricted action grid. We observe no supplier learning, consumer
population, real financial contracts, or disclosure policy. Broader
economic claims require repeated model sampling, new scenarios and model
families, plausible off-grid deviations, alternative strategic concepts,
and validation in a specified market. In finance, the natural next step
is to use verified payment-account or bank-fee schedules and realistic
customer profiles, together with costs of producing reliable
disclosures; those observations are not part of this study.

For research and deployment, the practical lesson is to evaluate AI
decision systems using the economic outcomes and incentives they create.
Fixed-menu accuracy, the magnitude of mistakes, the profitability of
seller actions, the stability of the strategic outcome, and the
incidence of gains are complementary measurements. Combining them makes
the limits of an agent's apparent capability visible and identifies the
assumptions needed before translating computational performance into
claims about consumer welfare.

The findings also suggest how a stronger evaluation should be organized.
The next study should repeat generations for each action-profile cell,
add independently generated scenarios and model families, and test finer
price grids or economically motivated off-grid deviations. A
common-interface factorial design could distinguish the role of verified
information from the role of deterministic calculation and final choice.
If the equilibrium result is to be used as a population claim, the study
should develop an inference procedure that accounts for payoff
estimation, equilibrium selection, and nonsmooth optimization rather
than relying on a percentile range alone.

An empirical application would add a different layer of evidence.
Researchers could collect a specified class of real contracts, document
how contractual totals are verified, and elicit or otherwise justify
customer usage and valuations. They would then need evidence about the
offers and information available to firms, as well as a realistic set of
actions that firms can change. Observed seller adaptation or a credible
institutional design would be needed to move from incentive-consistent
counterfactuals to claims about market adjustment. An H100 or another
faster accelerator could make additional inference runs feasible, but
computing capacity cannot substitute for these design and identification
choices.

The present contribution is therefore a disciplined separation of
objects that are often reported together. The calculator's advantage on
the fixed menu is real within the measured experiment; the disclosure
advantage under selected empirical equilibrium weights is a different
model-based comparison. The favorable conditional holdout result does
not remove the uncertainty of re-solving the game, and a
consumer-surplus gain is not automatically an equally large efficiency
gain. These distinctions allow the results to inform future economic
research without requiring the synthetic market to stand in for a
particular financial industry.

\FloatBarrier
\clearpage
\appendix
\renewcommand{\thetable}{\arabic{table}}
\renewcommand{\thefigure}{\arabic{figure}}

\hypertarget{scenario-generation-and-implementation}{%
\section{Scenario Generation and
Implementation}\label{scenario-generation-and-implementation}}

Table 10 collects the generator supports so that each transformation can
be interpreted economically. A representation changes the computational
path from these primitives to the same payment, not the primitives
themselves.

\begin{table}[H]
\caption{Scenario primitives and finite action supports.}
\label{tab:10}
\centering\small
\setlength{\tabcolsep}{3.2pt}
\renewcommand{\arraystretch}{1.12}
\begin{tabular}{@{}>{\raggedright\arraybackslash}p{108.82pt}>{\raggedright\arraybackslash}p{161.05pt}>{\raggedright\arraybackslash}p{165.40pt}@{}}
\toprule
\textbf{Quantity} & \textbf{Support or definition} & \textbf{Economic role} \\ \midrule
Horizon T & 3, 6, 12 months & Contract duration \\
Monthly usage u & 9, 13, 19, 27, 38 & Known demand \\
Base fee b & 8, 10, 12, 14 & Monthly fixed charge \\
Excess rate r & 0.20, 0.35, 0.50, 0.75 & Charge above allowance \\
Free allowance f & 5, 10, 15 units & Usage threshold \\
Monthly credit d & 1, 2, 3 & Conditional deduction \\
Credit threshold $\tau$ & 12, 20, 30 units & Credit eligibility \\
Opening fee s & 0, 3, 6, 9 & One-time payment \\
Monthly value v / T & 14, 19, 24, 29 & Given service valuation \\
Markup m & 0, 4, 8 & Seller price action \\
Representation k & 0, 1, 2 & Seller presentation action \\
\bottomrule
\end{tabular}
\vspace{3pt}\par\begin{minipage}{\linewidth}\footnotesize Artificial primitives, not fitted household or institution parameters. Full-horizon values equal monthly values times T. Sellers commit before a scenario is realized.\end{minipage}
\end{table}

The master seed is 20261026. Scenario-specific seeds enter a stable hash
that initializes separate pseudorandom generators. Each seller's
full-horizon service value equals its drawn monthly value multiplied by
\(T\). Resource cost is \(c_i=T(0.65b_i+0.30ur_i)+0.20s_i\). These are
experimental primitives, not estimates of institutional costs or
household preferences.

The Qwen3-8B revision is
\nolinkurl{b968826d9c46dd6066d109eabc6255188de91218}. Execution uses
PyTorch 2.8.0+cu128, CUDA 12.8, batch size 32, and a maximum context of
12,288 tokens, with prefix caching disabled. Context overflow raises an
error rather than silently truncating the contract. Reasoning
exhaustion, natural completion, raw answers, parsing failures, input and
output tokens, and timing are logged separately. The experiment does not
substitute corrected human answers for model failures.

The confirmation archive contains 1,224 atomic batches holding 33,048
decisions. Its JSONL export represents the same observations and is not
an additional sample. An independent CPU audit reconstructs inputs,
parses raw answers again, and recomputes economic outcomes for every
record. It also recomputes 66 policy-parameter CE bounds, all
leave-one-scenario-out estimates, and frozen-strategy holdout results.
The largest relevant numerical discrepancy is below \(2\times10^{-11}\).
For each panel, the audit independently re-solves the first 20 bootstrap
draws and checks the quantiles of all 1,000 stored draws; it does not
independently rerun all bootstrap optimizations.

\hypertarget{experimental-notebooks-and-data-provenance}{%
\section{Experimental Notebooks and Data
Provenance}\label{experimental-notebooks-and-data-provenance}}

The supplied notebooks represent three stages rather than four
independent experiments.
\nolinkurl{Randomized_Reasoning_Market_Colab_(1).ipynb} and
\nolinkurl{Randomized_Reasoning_Market_Colab_1027v1.ipynb} have
identical cell sources and saved outputs, although their complete file
hashes differ. Both are retained as original pilot records.
\nolinkurl{Randomized_Reasoning_Market_Colab_1027v3.ipynb} contains
diagnostics and calibration.
\nolinkurl{Reasoning_Market_Continue_Colab 1027.ipynb} implements the v4
continuation and defaults to the focused confirmation preset. The
executable source embedded in each distinct notebook is byte-identical
to the corresponding source bundle in its recorded results archive.

The central confirmation input is
\nolinkurl{reasoning_market_continue_v4_focused_vllm_results(1).zip},
with SHA256
\nolinkurl{7748bcf718ddbf582953c5e208229a71a6ada172b4c11cdf2caa557cb99b034f}.
The authoritative confirmation scope is specified by
\nolinkurl{plan.json}. Legacy scope flags inherited from development do
not establish that cross-model or off-grid confirmation was completed.
The v3 archive supplies development information to v4; its observations
are not pooled with the new confirmation scenarios.

The accompanying replication package preserves the four uploaded
notebooks, explains their order, and includes the tables, English
figure-generation code, model-free audit, and manuscript sources.

The expanded figure suite includes a separate program for rebuilding the
post hoc aggregates from the canonical records. The program verifies the
original archive hash, checks unique decision identifiers, and retains
exactly 31,104 core observations for the added loss and payoff-slice
diagnostics. It does not read the duplicate export or execute the
embedded GPU source. The compact aggregate files are sufficient to
redraw the expanded figures on a CPU; the original archive is needed to
repeat their construction from raw records. A separate CPU notebook
regenerates paper outputs from the audited summaries. It creates no new
model observations. Raw text and full logs remain in the original result
archives, whose names and hashes are documented. We checked notebook
structure and code-cell syntax and verified embedded-source
correspondence; the uploaded GPU experiments were not rerun while
preparing this manuscript.

\hypertarget{research-transparency}{%
\section{Research Transparency}\label{research-transparency}}

The study uses artificial contracts and model outputs rather than
personal financial records or human participants. Model responses,
deterministic parsing, and economic scoring are preserved separately.
The development trajectory, invalid outputs, negative budget findings,
and analyses marked post hoc remain visible. The financial
interpretation introduces no additional observations of institutions,
investors, or borrowers.

\hypertarget{construction-of-the-expanded-diagnostics}{%
\section{Construction of the Expanded
Diagnostics}\label{construction-of-the-expanded-diagnostics}}

The loss audit uses every canonical core record once. Each condition
contributes 48 scenarios, 81 joint actions per scenario, and two
positions per action. For response categories, validity and optimality
are crossed to obtain four mutually exclusive cells. Their counts sum to
7,776 within each condition. Invalid optimal choices are kept separate
because they arise from the outside-option fallback rather than
successful execution of the intended pipeline.

For the empirical loss distributions, the loss of each record is the
best available gross surplus, including the outside option, minus the
realized gross surplus. Equal weighting matches the static benchmark
used in the main comparison. The empirical cumulative distribution
assigns mass to every observed loss value; the percentile table uses
linear interpolation between ordered records. These summaries describe
the experimental menu. Scenario-cluster uncertainty would require
resampling whole scenarios and is not replaced by the apparent
smoothness of the plotted curves.

The fixed-opponent slice is constructed separately for each seller. For
seller 1, we retain records in which seller 2 has markup 4 and
complexity 1, then average seller 1's net profit for each of its nine
actions. For seller 2, the roles are reversed. Every seller-specific
action value averages 96 records, representing 48 scenarios and two
positions. The plotted value is the mean of these two seller-specific
averages. When the sellers choose the same action, the same market
record contributes to two different seller payoffs; it does not become
two independent market observations.

The transfer table loads the saved training joint distributions and
evaluates them directly against training and holdout payoff matrices. It
does not solve again to obtain a more favorable holdout strategy. The
same distributions are used to compute the largest positive joint
deviation gain. The new calculations reproduce the previously audited
diagonal cross-evaluation values for calculator and disclosure. This
agreement checks that the strategy indexing and weighting match the
original holdout analysis.

Table 11 implements the reweighting identity using the 48-scenario
static summary and each policy's consumer-worst empirical CE value at
\(\lambda=0.001\) and \(\kappa=0.05\). Uniform-menu net surplus equals
gross surplus minus \(0.001\) times the mean token proxy. The last
column subtracts the uniform value from the CE value within a policy;
its final row is the difference between the two changes. It is a post
hoc accounting calculation, not a new equilibrium solve or a causal
estimate.

\begin{table}[H]
\caption{The welfare comparison under uniform and equilibrium offer weights.}
\label{tab:11}
\centering\small
\setlength{\tabcolsep}{3.2pt}
\renewcommand{\arraystretch}{1.12}
\begin{tabular}{@{}>{\raggedright\arraybackslash}p{158.68pt}>{\centering\arraybackslash}p{85.77pt}>{\centering\arraybackslash}p{90.06pt}>{\centering\arraybackslash}p{94.35pt}@{}}
\toprule
\textbf{Buyer pipeline} & \textbf{Uniform net CS} & \textbf{CE lower net CS} & \textbf{Reweighting gap} \\ \midrule
Calculator 256 & 31.43 & 32.92 & +1.48 \\
Disclosure 256 & 26.47 & 53.06 & +26.58 \\
Disclosure $-$ calculator & -4.96 & +20.14 & +25.10 \\
\bottomrule
\end{tabular}
\vspace{3pt}\par\begin{minipage}{\linewidth}\footnotesize Post hoc accounting on the same 48 scenarios, at $\lambda$ = 0.001 and $\kappa$ = 0.05. Each CE column uses that policy’s own consumer-worst empirical CE. The final column subtracts uniform from CE surplus within policy; the final row is the difference of these gaps. These are simulation units and are not causal effects.\end{minipage}
\end{table}

Figure 3 presents conditional holdout and re-solved sensitivity
intervals; Figure 9 reports within-game CE ranges. These objects are not
interchangeable. All added diagnostics are post hoc and use the existing
33,048 confirmation decisions.

\FloatBarrier
\begingroup\small\singlespacing\setlength{\bibsep}{2pt}
\bibliographystyle{elsarticle-harv}
\bibliography{references}

@article{Aumann1974,
 author={Aumann, Robert J.}, title={Subjectivity and Correlation in Randomized Strategies},
 journal={Journal of Mathematical Economics}, year={1974}, volume={1}, number={1}, pages={67--96}, doi={10.1016/0304-4068(74)90037-8}}

@article{Bertrand2011,
 author={Bertrand, Marianne and Morse, Adair}, title={Information Disclosure, Cognitive Biases, and Payday Borrowing},
 journal={The Journal of Finance}, year={2011}, volume={66}, number={6}, pages={1865--1893}, doi={10.1111/j.1540-6261.2011.01698.x}}

@techreport{Beshears2009,
 author={Beshears, John and Choi, James J. and Laibson, David and Madrian, Brigitte C.},
 title={How Does Simplified Disclosure Affect Individuals' Mutual Fund Choices?},
 institution={National Bureau of Economic Research}, number={14859}, year={2009}, url={https://www.nber.org/papers/w14859}}

@misc{Buecheler2026,
 author={Buecheler, F. and Berbeglia, G. and Guo, M.},
 title={A Cue's Hidden Price: Auditing Seller Influence on {AI} Shoppers},
 year={2026}, howpublished={arXiv:2610.25843}, note={Preprint}, url={https://arxiv.org/abs/2610.25843}}

@article{Calvano2020,
 author={Calvano, Emilio and Calzolari, Giacomo and Denicol{\`o}, Vincenzo and Pastorello, Sergio},
 title={Artificial Intelligence, Algorithmic Pricing, and Collusion},
 journal={American Economic Review}, year={2020}, volume={110}, number={10}, pages={3267--3297}, doi={10.1257/aer.20190623}}

@article{Campbell2011,
 author={Campbell, John Y. and Jackson, Howell E. and Madrian, Brigitte C. and Tufano, Peter},
 title={Consumer Financial Protection}, journal={Journal of Economic Perspectives}, year={2011}, volume={25}, number={1}, pages={91--114}, doi={10.1257/jep.25.1.91}}

@article{Carlin2009,
 author={Carlin, Bruce I.}, title={Strategic Price Complexity in Retail Financial Markets},
 journal={Journal of Financial Economics}, year={2009}, volume={91}, number={3}, pages={278--287},
 url={https://www.sciencedirect.com/science/article/abs/pii/S0304405X08002092}}

@article{Choi2010,
 author={Choi, James J. and Laibson, David and Madrian, Brigitte C.},
 title={Why Does the Law of One Price Fail? An Experiment on Index Mutual Funds},
 journal={The Review of Financial Studies}, year={2010}, volume={23}, number={4}, pages={1405--1432}, doi={10.1093/rfs/hhp097}}

@article{Ellison2009,
 author={Ellison, Glenn and Ellison, Sara Fisher}, title={Search, Obfuscation, and Price Elasticities on the Internet},
 journal={Econometrica}, year={2009}, volume={77}, number={2}, pages={427--452}, doi={10.3982/ECTA5708}}

@article{Fang2019,
 author={Fang, Zheng and Santos, Andres}, title={Inference on Directionally Differentiable Functions},
 journal={The Review of Economic Studies}, year={2019}, volume={86}, number={1}, pages={377--412}, doi={10.1093/restud/rdy049}}

@article{Gabaix2006,
 author={Gabaix, Xavier and Laibson, David}, title={Shrouded Attributes, Consumer Myopia, and Information Suppression in Competitive Markets},
 journal={The Quarterly Journal of Economics}, year={2006}, volume={121}, number={2}, pages={505--540},
 url={https://academic.oup.com/qje/article-abstract/121/2/505/1884013}}

@misc{Horton2026,
 author={Horton, John J. and Filippas, Apostolos and Manning, Benjamin S.},
 title={Large Language Models as Simulated Economic Agents: What Can We Learn from {Homo silicus}?},
 year={2026}, howpublished={arXiv:2301.07543v2}, note={Preprint; first version 2023}, url={https://arxiv.org/abs/2301.07543v2}}

@inproceedings{Kwon2023,
 author={Kwon, Woosuk and Li, Zhuohan and Zhuang, Siyuan and others},
 title={Efficient Memory Management for Large Language Model Serving with {PagedAttention}},
 booktitle={Proceedings of the 29th Symposium on Operating Systems Principles}, year={2023}, pages={611--626}, doi={10.1145/3600006.3613165}}

@article{Simon1955,
 author={Simon, Herbert A.}, title={A Behavioral Model of Rational Choice},
 journal={The Quarterly Journal of Economics}, year={1955}, volume={69}, number={1}, pages={99--118}, url={https://www.jstor.org/stable/1884852}}

@article{Sims2003,
 author={Sims, Christopher A.}, title={Implications of Rational Inattention},
 journal={Journal of Monetary Economics}, year={2003}, volume={50}, number={3}, pages={665--690}, doi={10.1016/S0304-3932(03)00029-1}}

@misc{Wadi2026,
 author={Wadi, D. and Ma, Y.}, title={Shopping by Algorithm: How Agentic {AI} Deploys Human Heuristics as a Surrogate Consumer},
 year={2026}, howpublished={arXiv:2609.28372v2}, note={Preprint}, url={https://arxiv.org/abs/2609.28372v2}}

@misc{Yang2025,
 author={Yang, An and Li, Anfeng and Yang, Baosong and others}, title={{Qwen3} Technical Report},
 year={2025}, howpublished={arXiv:2505.09388}, url={https://arxiv.org/abs/2505.09388}}

@article{Chetty2009,
 author={Chetty, Raj and Looney, Adam and Kroft, Kory},
 title={Salience and Taxation: Theory and Evidence},
 journal={American Economic Review}, year={2009}, volume={99}, number={4}, pages={1145--1177}, doi={10.1257/aer.99.4.1145}}

@article{Kleinberg2015,
 author={Kleinberg, Jon and Ludwig, Jens and Mullainathan, Sendhil and Obermeyer, Ziad},
 title={Prediction Policy Problems},
 journal={American Economic Review}, year={2015}, volume={105}, number={5}, pages={491--495}, doi={10.1257/aer.p20151023}}

@inproceedings{Saig2024,
 author={Saig, Eden and Einav, Ohad and Talgam-Cohen, Inbal},
 title={Incentivizing Quality Text Generation via Statistical Contracts},
 booktitle={Advances in Neural Information Processing Systems}, volume={37}, year={2024}, pages={51196--51222},
 doi={10.52202/079017-1620}, url={https://proceedings.neurips.cc/paper_files/paper/2024/hash/5b93ce41ac6de2bf9aca7e4ba5ba01d5-Abstract-Conference.html}}

@article{Brown2023,
 author={Brown, Zach Y. and MacKay, Alexander}, title={Competition in Pricing Algorithms},
 journal={American Economic Journal: Microeconomics}, year={2023}, volume={15}, number={2}, pages={109--156}, doi={10.1257/mic.20210158}}

@inproceedings{Hardt2016,
 author={Hardt, Moritz and Megiddo, Nimrod and Papadimitriou, Christos H. and Wootters, Mary},
 title={Strategic Classification},
 booktitle={Proceedings of the 2016 ACM Conference on Innovations in Theoretical Computer Science},
 year={2016}, pages={111--122}, doi={10.1145/2840728.2840730}}

@inproceedings{Perdomo2020,
 author={Perdomo, Juan and Zrnic, Tijana and Mendler-D{\"u}nner, Celestine and Hardt, Moritz},
 title={Performative Prediction},
 booktitle={Proceedings of the 37th International Conference on Machine Learning},
 series={Proceedings of Machine Learning Research}, volume={119}, pages={7599--7609}, year={2020},
 url={https://proceedings.mlr.press/v119/perdomo20a.html}}
\endgroup

\end{document}